\documentclass[conference]{IEEEtran}
\IEEEoverridecommandlockouts
\usepackage{cite}
\usepackage{amsmath,amssymb,amsfonts}
\usepackage{algorithmic}
\usepackage{graphicx}
\usepackage{textcomp}
\usepackage{xcolor}
\usepackage{booktabs}
\usepackage{multirow}
\usepackage{subfigure}
\usepackage{enumitem}
\usepackage{bm}
\usepackage{url}       
\usepackage{ulem} 
\usepackage{array}
\usepackage{hyperref}
\hypersetup{
	pdfborder={0 0 0},     
	colorlinks=true,        
	linkcolor=black,
	citecolor=black,
	urlcolor=black
}

\newcommand{\hide}[1]{} 

\renewcommand{\IEEEauthorrefmark}[1]{$^{\text{#1}}$}
\def\BibTeX{{\rm B\kern-.05em{\sc i\kern-.025em b}\kern-.08em
		T\kern-.1667em\lower.7ex\hbox{E}\kern-.125emX}}

\newcommand{\sRC}{AGR~}
\begin{document}
	\title{Enhancing Group Recommendation with Memory-Augmented Reasoning in LLM Agent}
	\author{
		\IEEEauthorblockN{
			Qimeng Niu\IEEEauthorrefmark{1},
			Bowen Hao\IEEEauthorrefmark{1}\textsuperscript{\dag},
			Zixuan Zhang\IEEEauthorrefmark{1},
			Shuyu Qu\IEEEauthorrefmark{1},
			Hongzhi Yin\IEEEauthorrefmark{2}
		}
		\IEEEauthorblockA{\IEEEauthorrefmark{1}
			\textit{School of Management}, \textit{Capital Normal University}, \textit{Beijing, China}\\
			\hide{Beijing, China \\}
			\{1232905021, 6974, 1222905049, 1232905022\}@cnu.edu.cn\\
			\textsuperscript{\dag} Corresponding author: 6974@cnu.edu.cn.
		}
		\IEEEauthorblockA{\IEEEauthorrefmark{2}
			\textit{School of Electrical Engineering and Computer Science}, \textit{The University of Queensland} \\
			\hide{Brisbane, Australia \\}
			h.yin1@uq.edu.cn
		}
	}
	\hide{\author{
			\IEEEauthorblockN{1\textsuperscript{st} Qimeng Niu}
			\IEEEauthorblockA{\textit{School of Management} \\
				\textit{Capital Normal University}\\
				Beijing, China \\
				1232905021@cnu.edu.cn}
			\and
			\IEEEauthorblockN{2\textsuperscript{nd} Bowen Hao}
			\IEEEauthorblockA{\textit{School of Management} \\
				\textit{Capital Normal University}\\
				Beijing, China \\
				6974@cnu.edu.cn}
			\and
			\IEEEauthorblockN{3\textsuperscript{rd} Zixuan Zhang}
			\IEEEauthorblockA{\textit{School of Management} \\
				\textit{Capital Normal University}\\
				Beijing, China \\
				1232905049@cnu.edu.cn}
			\and
			\IEEEauthorblockN{4\textsuperscript{th} Shuyu Qu}
			\IEEEauthorblockA{\textit{School of Management} \\
				\textit{Capital Normal University}\\
				Beijing, China \\
				1232905022@cnu.edu.cn}
			\and
			\IEEEauthorblockN{5\textsuperscript{th} Hongzhi Yin}
			\IEEEauthorblockA{\textit{the School of Electrical Engineering and Computer Science} \\
				\textit{The University of Queensland}\\
				Brisbane, Australia \\
				h.yin1@uq.edu.au}}
	}
	
	\maketitle
	\vspace*{-0.35in}
	\begin{abstract}
		The core challenge in group recommendation lies in modeling the dynamic evolution of user preferences and explaining the consensus formation process. Existing Large Language Model (LLM)-based methods, despite improved interpretability, treat interaction history as fixed text, ignoring the natural evolution of group/user preferences over time, and lacking explicit modeling of the complex group decision‑making process. To address these issues, we propose AGR, a LLM‑based agent, which consists of a Memory Module and a Reasoning Module. The Memory Module employs a token‑based hash table to dynamically manage the historical interactions of groups and users. This design supports fundamental operations including insertion, updating, retrieval, forgetting of irrelevant records, and summarization of evolving group and user profiles for efficiently tracking.
		Based on these retrieved dynamic profiles, the Reasoning Module then performs a multi‑step reasoning process including Group Interests Collection, Group Consensus Refinement, Multi‑dimensional Evaluation and Explainable Recommendation Generation, thereby moving beyond black‑box inference to deliver fully interpretable recommendations. In practice, we adopt the Reinforcement Fine‑Tuning (RFT) paradigm, where we first use Supervised Fine‑Tuning (SFT) to equip the model with basic capabilities for invoking the Memory and Reasoning modules, and then employ Group Relative Policy Optimization (GRPO) to enhance its autonomous ability to coordinate these modules. Experiments on LastFM and Douban datasets demonstrate that AGR significantly outperforms existing state‑of‑the‑art methods in both recommendation accuracy and explainability. Our model is open‑sourced at https://huggingface.co/niuqimeng/AGR.
	\end{abstract}
	
	
	
	\section{Introduction}
Recommender systems have become essential tools for mitigating information overload. As an important branch, group recommendation has garnered considerable attention due to its wide application in scenarios such as family viewing, friend gatherings, and team travel~\cite{DBLP:conf/icde/YinW0LYZ19}. The principal technical challenge in group recommendation resides in the derivation of fair, rational, and explainable collective decisions from a set of diverse and potentially conflicting individual preferences.

Traditional group recommendation methods are broadly categorized into score aggregation and profile aggregation methodologies. Score aggregation methods first estimate individual preference scores for candidate items and subsequently aggregate these scores via rule-based strategies (e.g., average~\cite{DBLP:conf/recsys/BaltrunasMR10}, least misery~\cite{DBLP:journals/pvldb/Amer-YahiaRCDY09}, or maximum satisfaction~\cite{DBLP:series/sci/BorattoC11}) to produce a final group recommendation. However, these models rely on heuristic rules and struggle to flexibly model group decision processes. In contrast, profile aggregation methods construct group representations by fusing member profiles and then perform recommendations~\cite{DBLP:conf/icde/YinW0LYZ19}. Although this paradigm more closely approximates the intrinsic nature of the group decision process, the internal mechanisms of profile fusion frequently operate as an opaque ``black box", resulting in a significant deficit in model explainability.

\begin{figure}[t]
	\centering
	\includegraphics[width=0.47\textwidth]{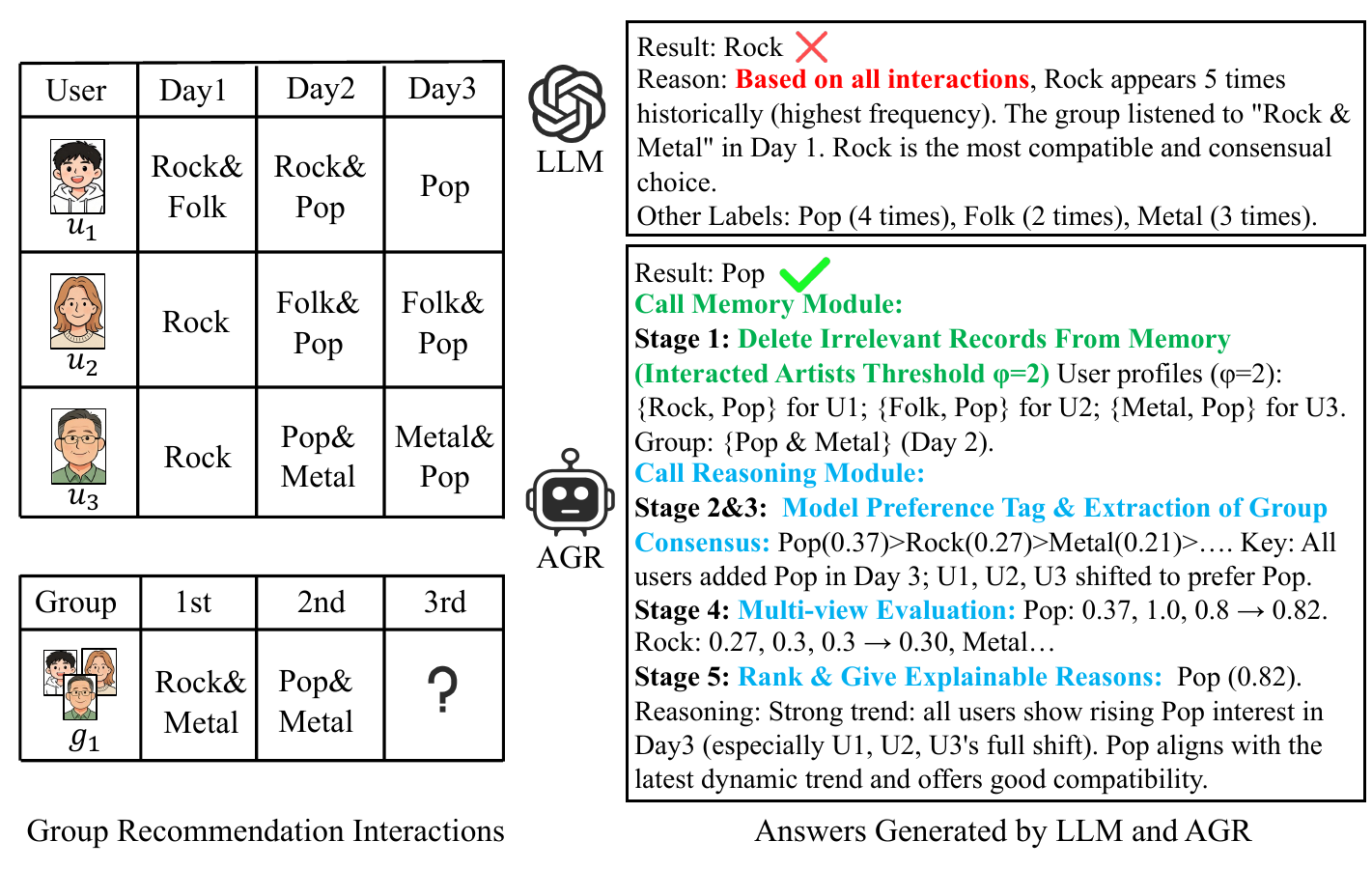}
	\caption{Recommendation results for group $g_1 = \{u_1, u_2, u_3\}$ on four candidate music genres (Rock, Metal, Pop, Folk) by Vanilla LLM and AGR. "1st", "2nd", "3rd" represent the group's first, second, and third gathering scenarios.}
	\label{fig:example}
\end{figure}

Recently, the advent of Large Language Models (LLMs) has opened new avenues for enhancing the explainability of group recommendation~\cite{DBLP:conf/intrs/LubosTLGHWF25}. Existing methods~\cite{DBLP:conf/ictai/LubosFGL25,DBLP:conf/um/LubosFGLHWF25} typically feed group and member interactions directly into the vanilla LLM to generate both recommendations and decision rationales. However, these LLM-based methods still suffer from the following two challenges. First, they \textbf{ignore the temporal evolution of individual preferences within a group} (as shown in Fig.~\ref{fig:example}, three members $u_1$, $u_2$, and $u_3$ show different interests over three days). By treating all historical interactions with equal weight, existing methods fail to model dynamic preference shifts. As a result, they introduce a bias that distorts user profiles and further distorts the group decision-making process.
Second, group decision-making involves negotiation, compromise, and consensus-building among members. Current LLM-based methods rely on the model's implicit internal reasoning to infer group preferences, \textbf{failing to explicitly model the complex group decision-making process}, leading to superficial explanations (e.g., "because most people like it") that do not reveal the true consensus logic. Consequently, there is a clear need for a tailored, explicit multi-stage reasoning process for group recommendation.

Inspired by the autonomous reasoning and memory management capabilities of LLM agents~\cite{DBLP:journals/tmlr/ZhangGYYZTZLXLZCZFWHVLW26}, we propose AGR, the first LLM \underline{A}gent framework for \underline{g}roup \underline{r}ecommendation, which formulates group recommendation as a sequential decision-making task driven by an agent equipped with memory and reasoning capabilities. Specifically, AGR consists of a Memory Module and a Reasoning Module. The Memory Module employs a token-based hash table to dynamically manage the historical interactions of groups and users. This design supports fundamental operations including insertion, updating, retrieval, forgetting of irrelevant records, and summarization of evolving group and user profiles for efficiently tracking. The Reasoning Module then performs a multi-step reasoning process based on the group profiles, user profiles, and item attributes retrieved by the Memory Module. This process includes Group Interests Collection, Group Consensus Refinement, Multi-dimensional Evaluation, and Explainable Recommendation Generation, thereby moving beyond black-box inference to deliver fully interpretable recommendations. As shown in Fig.~\ref{fig:example}, compared to the vanilla LLM, AGR is able to capture updates of dynamic group/user profiles and show explicit multi-stage reasoning capabilities, thus leading to recommendations that better align with the group's latest preferences. To effectively train the agent to autonomously invoke these modules, we employ a Reinforcement Fine-tuning (RFT) paradigm~\cite{DBLP:conf/nips/SchickDDRLHZCS23} which consists of two steps: Supervised Fine-Tuning (SFT) to establish foundational task comprehension, followed by Group Relative Policy Optimization (GRPO) to reinforce the agent's strategic coordination of memory retrieval and multi-step reasoning. Experimental results on the LastFM and Douban datasets show that AGR achieves significant improvements over state-of-the-art baselines in both top-$k$ recommendation performance and the quality of generated explanations.
Our contributions can be summarized as follows:
\begin{itemize}[leftmargin=1pt]
	\item We propose AGR, the first agent equipped with memory-augmented reasoning capabilities that dynamically manages the profiles of the users and groups to address the group recommendation task.
	\item We design a Memory Module to handle evolving group/user profiles and a Reasoning Module that explicitly models group decision-making to enhance transparency and interpretability.
	\item Experiments on two benchmark datasets show that AGR outperforms existing methods in both accuracy and explainability.
\end{itemize}
	\section{Related Work}
\subsection{Group Recommendation}
Existing group recommendation approaches mainly follow two paradigms: traditional aggregation and LLM-based methods. Traditional aggregation operates either through score aggregation, where member ratings are combined via fixed functions (e.g., averaging~\cite{DBLP:conf/recsys/BaltrunasMR10}, least misery~\cite{DBLP:journals/pvldb/Amer-YahiaRCDY09}, or maximum satisfaction~\cite{DBLP:series/sci/BorattoC11}) that can lead to rigid outcomes, or through profile aggregation, which integrates user profiles using techniques such as probabilistic generative models~\cite{DBLP:conf/kdd/YuanCL14} or attention mechanisms~\cite{DBLP:conf/icde/YinW0LYZ19} but often suffers from low interpretability. While LLM-based methods leverage semantic understanding of group histories~\cite{DBLP:conf/ictai/LubosFGL25,DBLP:conf/um/LubosFGLHWF25}, they treat these histories as static and overlook temporal preference evolution.

\subsection{Memory-based Recommendation}
Memory mechanisms are crucial for capturing dynamic user preferences. Early works used RNNs (e.g., LSTM~\cite{DBLP:journals/neco/HochreiterS97}, GRU~\cite{DBLP:conf/ssst/ChoMBB14}) to model interaction sequences but faced vanishing gradient issues in long sequences. Subsequent works introduced explicit memory modules to store user historical interactions for long-term interest modeling\hide{~\cite{DBLP:conf/sigir/HuangZDWC18}}. Recently, LLM-based memory-augmented systems (e.g., MARM~\cite{DBLP:journals/corr/abs-2411-09425}, LMN~\cite{DBLP:conf/www/LuCZCXXZW25}, MemoCRS~\cite{DBLP:conf/cikm/XiLL0T0024}, MR.Rec~\cite{DBLP:journals/corr/abs-2510-14629}) retrieve past interactions from memory to boost performance but rely on heuristic operations (insertion, deletion, updating, summarization) and neglect temporal profile evolution. Moreover, they are primarily designed for personalized recommendation scenarios and are not applicable to group recommendation.

\subsection{Reasoning-based Recommendation}
Reasoning-based recommendation seeks to enhance model explainability by incorporating explicit decision processes. Symbolic methods (e.g., KGAT~\cite{DBLP:conf/kdd/Wang00LC19}, RuleRec~\cite{DBLP:conf/www/MaZCJWLMR19}) depend on handcrafted rules, limiting their ability to model complex dynamics or generate natural-language explanations. Neural models (e.g., SASRec~\cite{DBLP:conf/icdm/KangM18}) lack semantic interpretability and resort to post-hoc attribution. LLM-based reasoning (e.g., CoT4Rec~\cite{DBLP:conf/aaai/YueYZSLW25}, RecExplainer~\cite{DBLP:conf/kdd/LeiLYHL024}) offers explanations but struggles with temporal preference evolution and lacks explicit modeling of intricate group decision-making.
	\section{Preliminary}

\noindent \textbf{Problem Definition:}
Let $G = \{g_1, \ldots, g_{|G|}\}$ denote the set of groups, $U = \{u_1, \ldots, u_{|U|}\}$ the set of users, and $I = \{i_1, \ldots, i_{|I|}\}$ the set of items. To capture the dynamic evolution of group/user interests and the group decision process, we formalize group recommendation as a sequential decision-making task. Formally, for a given group $g_j = \{u_1^j, u_2^j, \ldots\} \in G$ with historical interaction item set $\mathcal{H}_{t-1}^{g_j} = \{ s_1^{g_j}, s_2^{g_j}, \ldots, s_{t-1}^{g_j}\}$ prior to time $t$, at time $t$, the LLM Agent $f_{\theta}(\cdot)$ is required to recommend top-$k$ items from the candidate set $\mathcal{C} = I - \mathcal{H}_{t-1}^{g_j}$ to the group, i.e., $o_j = \mathbf{\mathit{f}}_{\theta}(\mathcal{R}(\mathcal{M}( g_j, \mathcal{H}_{t-1}^{g_j} )))$, where $o_j$ denotes the top-$k$ recommendations and their explanations, $\mathcal{M}(\cdot)$ is the Memory Module responsible for managing  group/user profiles; $\mathcal{R}(\cdot)$ is the Reasoning Module which leverages the retrieved profiles alongside item attributes from the Memory Module to perform a structured multi-stage reasoning process, and the LLM Agent $f_{\theta}(\cdot)$  aims to generate the final recommendation and corresponding explanations, based on intermediate results from the Memory and Reasoning Modules.

	\section{Methodology}
\begin{figure*}[ht]
	\centering
	\includegraphics[width=0.89\textwidth]{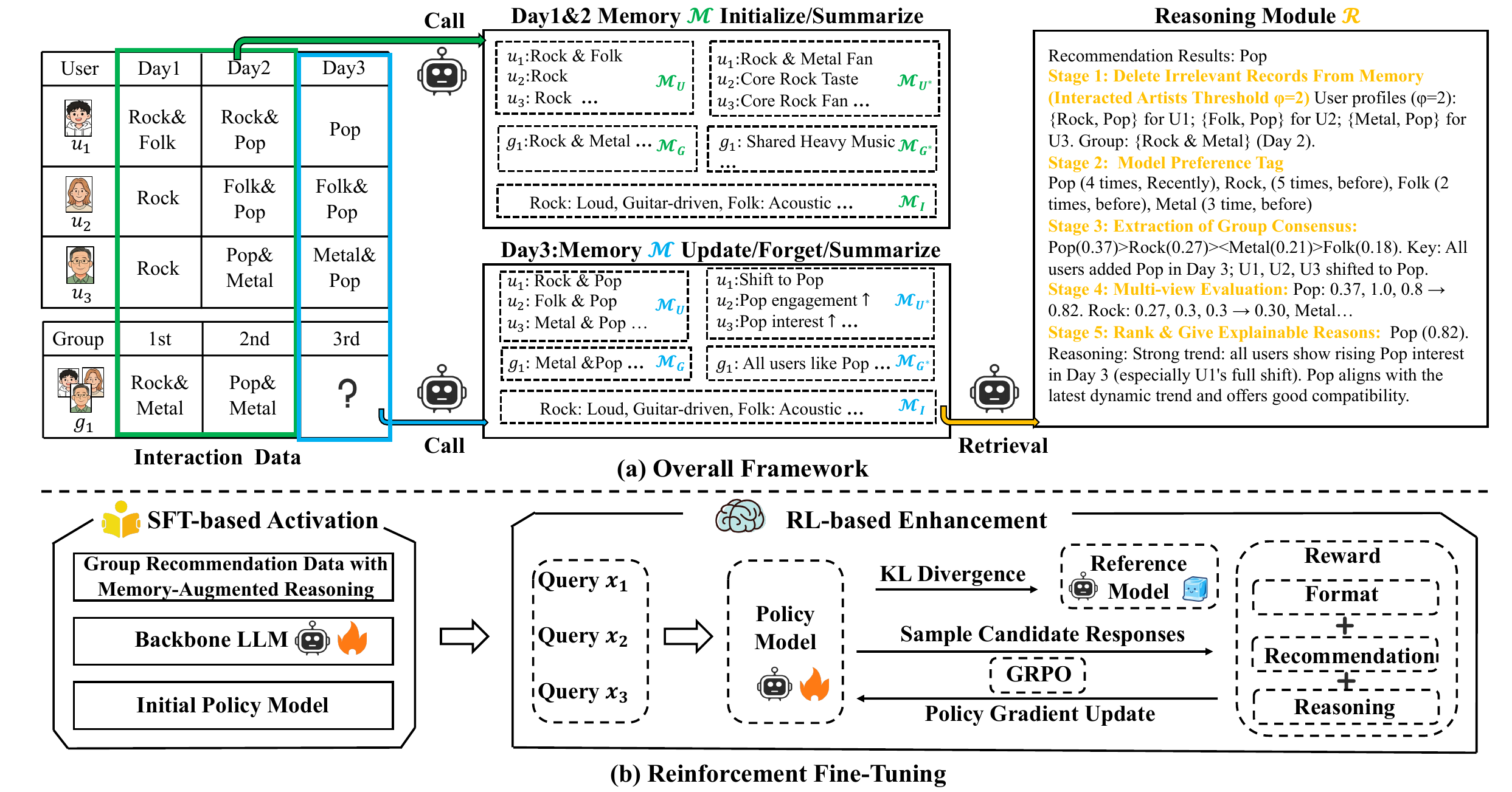}
	\caption{The AGR framework. (a) Workflow for AGR to invoke the Memory and Reasoning Modules during group recommendation. (b) RFT training pipeline.}
	\label{fig:framework}
\end{figure*}

We propose AGR, a framework that consists of a Memory Module and a Reasoning Module. The Memory Module employs a token-based hash table to dynamically manage the historical interactions of groups and users. The Reasoning Module then performs a multi-step reasoning process based on the group profiles, user profiles, and item attributes retrieved from the Memory Module, thereby moving beyond ``black-box" inference to deliver fully interpretable recommendations. In practice, we perform RFT~\cite{DBLP:conf/nips/SchickDDRLHZCS23} to enable AGR to autonomously invoke the Memory and Reasoning Modules. Fig.~\ref{fig:framework} illustrates the overall framework.

\hide{
	\textcolor{green}{\subsection{Dynamic Trigger Mechanism}
		To address the inefficiency and rigidity of the fixed heuristic pipeline, AGR adopts a dynamic trigger mechanism to adaptively activate memory retrieval and reasoning modules. To avoid frequent updates caused by unstable user/group profiles, we set an interaction count threshold $\theta=7$. Before the cumulative number of interactions reaches the threshold, the system only accumulates behaviors without triggering memory updates and forgetting. After exceeding the interaction threshold, the system triggers full memory retrieval and multi-stage reasoning if either of the following conditions is met:
		1) The cosine similarity $\text{sim}$ between the current group/user profile $P_{\text{cur}}$ and the previous time-step profile $P_{\text{prev}}$ satisfies $\text{sim}<\theta_{\text{sim}}$ ($\theta_{\text{sim}}<0.4$), indicating a significant preference shift;
		2) The time interval since the last activation exceeds $\tau=3$ time steps to ensure long-term evolution is captured.
		Otherwise, the system performs a lightweight update: the memory module inserts and updates interaction records normally, while the reasoning module skips consensus refinement and multi-dimensional evaluation, and directly generates recommendations based on existing profiles and similarity ranking, maintaining efficiency and recommendation coherence.}
}

\subsection{Memory Module}
\label{sec:memory}
The Memory Module $\mathcal{M}$ is the core component for modeling dynamic user and group interests. Built upon a token-based hash table, $\mathcal{M}$ consists of the following subunits:
\begin{itemize}[leftmargin=10pt]
	\item $\mathcal{M}_G$: Stores time-ordered group interactions using group ID as the key and a natural language description of group interactions as the value.
	\item $\mathcal{M}_U$: Stores time-ordered user interactions using user ID as the key and a natural language description of user interactions as the value.
	\item $\mathcal{M}_I$: Stores descriptions of items using item ID as the key and a natural language description of item attributes as the value.
	\item $\mathcal{M}_{G^*}$: Stores dynamically generated group profiles derived from $\mathcal{M}_G$, using the group ID as the key and a natural language description of the group profile as the value.
	\item $\mathcal{M}_{U^*}$: Stores dynamically generated user profiles derived from $\mathcal{M}_U$, using the user ID as the key and a natural language description of the user profile as the value.
\end{itemize}

\begin{figure}[h!]
	\centering
	\includegraphics[width=0.46\textwidth]{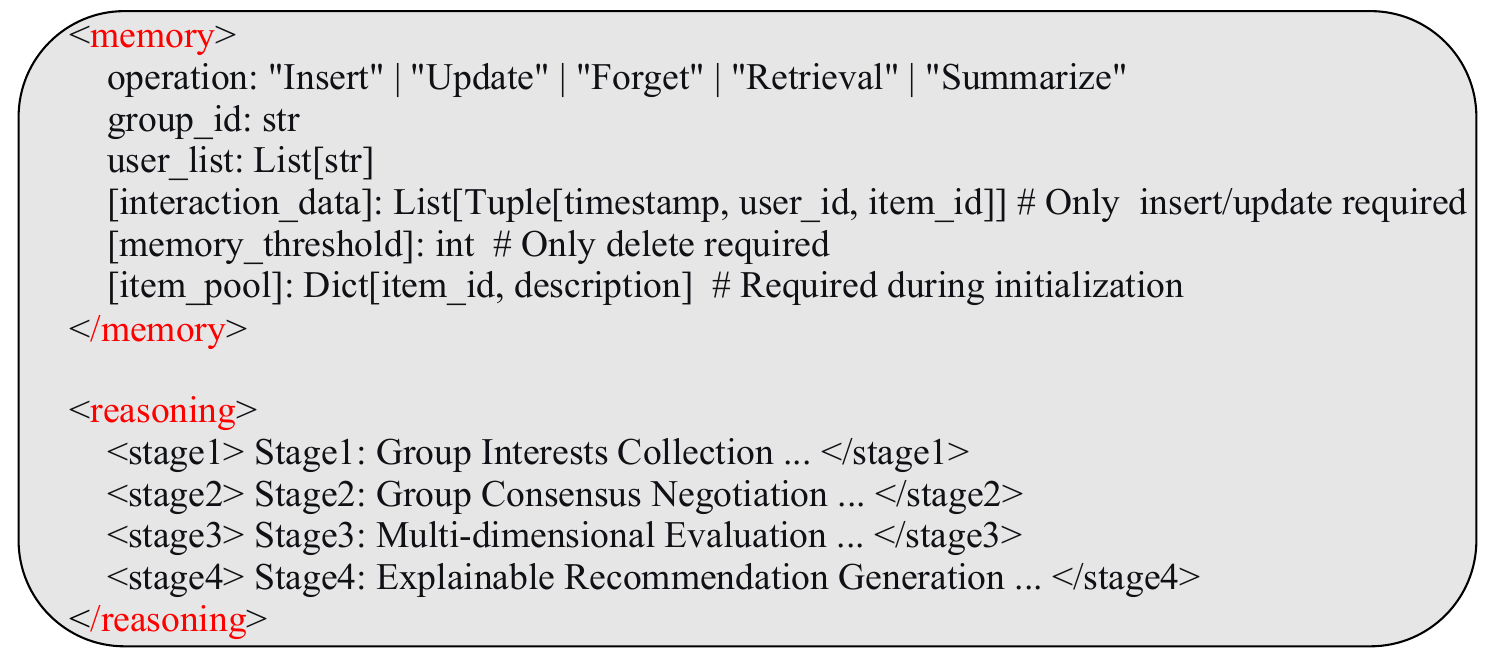}
	\caption{Interface for the Memory and Reasoning Modules.}
	\label{fig:Memory_Reasoning_Interface}
\end{figure}

To efficiently model and query the dynamic interests of groups and users, inspired by Toolformer~\cite{DBLP:conf/nips/SchickDDRLHZCS23}, we endow $\mathcal{M}$ with five fundamental operations and implement them as functions that can be invoked directly by LLMs. The interface for these functions is depicted in Fig.~\ref{fig:Memory_Reasoning_Interface}, and the details are as follows:

\begin{itemize}[leftmargin=10pt]
	\item \textbf{Insertion}: appends group and user interactions to $\mathcal{M}_G$ and $\mathcal{M}_U$ as sequential key-value pairs. Since the hash table allows direct access by ID and appending, it takes $\mathcal{O}(1)$ time.
	\item \textbf{Updating}: when new interactions occur from $t-1$ to $t$, append them to the end of the corresponding user or group's memory sequence. This triggers an updated summary of the group/user profile and also takes $\mathcal{O}(1)$ time.
	\item \textbf{Forgetting}: To prevent memory overload, once the number of historical interactions surpasses a predefined threshold $\varphi$, the proposed mechanism discards items based on the relevance between the item and the current group profile, rather than simply removing the oldest interacted items. Specifically, the process begins by scanning items from the earliest record and computing the cosine similarity between each item and the current group profile (derived from the summary operation). The item with the lowest similarity score is selected for removal.
	This approach preserves long-term stable preferences essential for group decision-making, with a time complexity of $\mathcal{O}(\varphi)$.
	\item \textbf{Summarization:} After initialization, updating, or forgetting operations, we generate group and user profile summaries ($\mathcal{M}_{G^*}$, $\mathcal{M}_{U^*}$) based on $\mathcal{M}_G$ and $\mathcal{M}_U$.  This operation runs in $\mathcal{O}(n) + {\rm LLM\_cost}(n)$ time, where $n$ denotes the number of historical interaction items associated with a given group or user, and $ {\rm LLM\_cost}(n)$ represents the generation time incurred by the LLM. Notably, any modification to the historical records (i.e., insertion, updating, or forgetting) triggers a full summary regeneration.

	\item \textbf{Retrieval}: The retrieval phase includes two components: Historical Interaction Lookup, which relies on a hash-based process with $\mathcal{O}(n)$ time complexity for $n$ groups or members, and a Ranking Retrieval process composed of two stages. First, Coarse Semantic Filtering embeds groups and items using Qwen3~\cite{DBLP:journals/corr/abs-2505-09388} and applies the FAISS-HNSW indexing algorithm~\cite{8594636}, which offers $\mathcal{O}({\rm log}(|I|))$ complexity to narrow down the candidate set to the top-$K$ ($K \ll |I|$ ) items. Second, Fine-grained LLM Filtering sends profile descriptions to an LLM (e.g., DeepSeek-R1) with the prompt: \textit{``Based on group and user profiles, select items highly relevant to the group's interests from the Candidate Item List $\mathcal{C}$ =  […] "}.  Matching all items takes $\mathcal{O}(|K|)$ time. Consequently, the total time complexity is $\mathcal{O}({\rm log}(|I|))$ + $\mathcal{O}(|K|)$.
	
\end{itemize}

\subsection{Memory-Augmented Reasoning Module}
\begin{figure}[h!]
	\centering
	\includegraphics[width=0.48\textwidth]{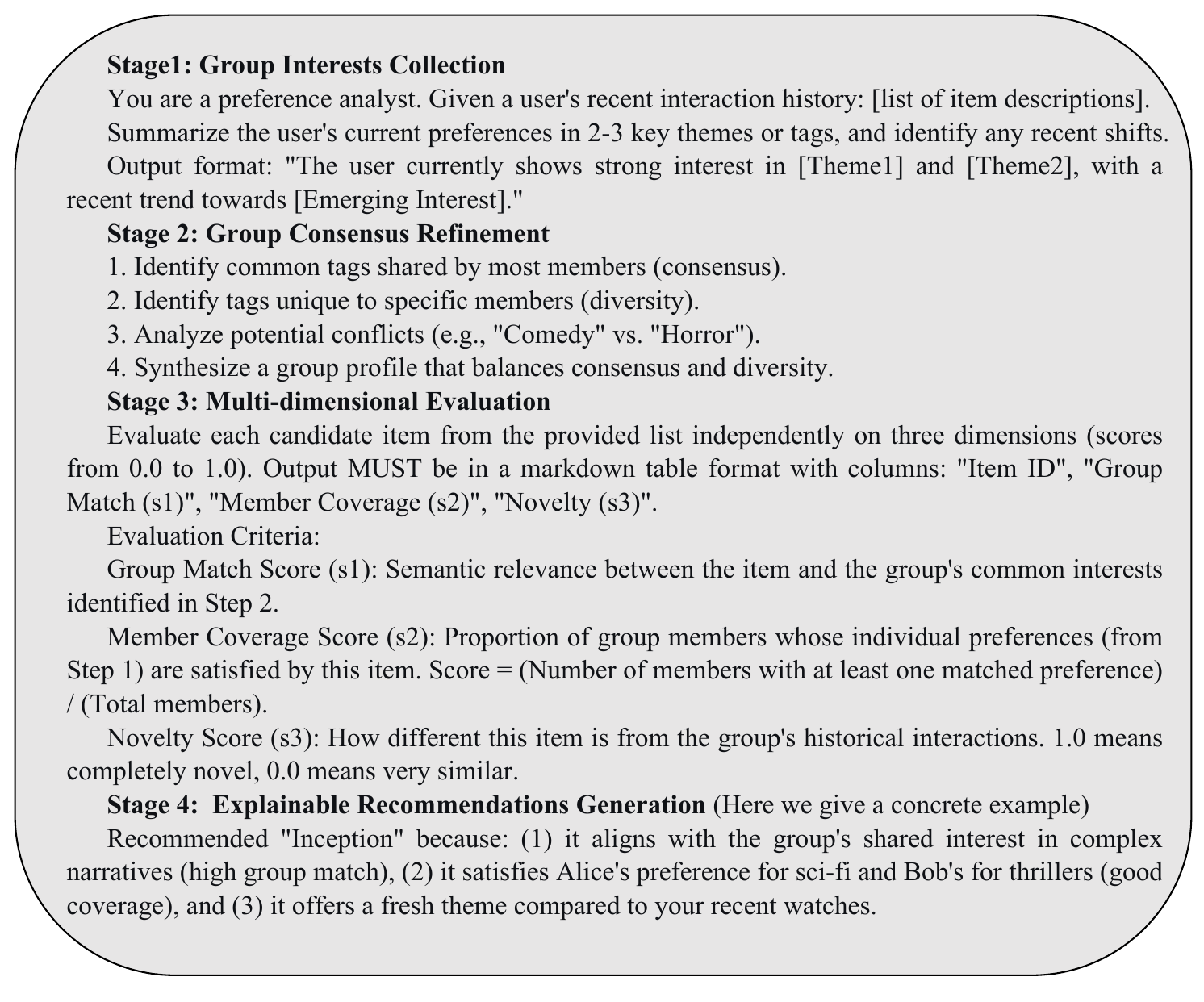}
	\caption{Prompt templates for the Reasoning Module.}
	\label{fig:prompts}
\end{figure}

While LLMs exhibit strong general reasoning capabilities, they are not tailored for group recommendation and cannot effectively model the group decision-making process. To bridge this gap, we introduce a Reasoning Module that performs a multi-step reasoning process based on the retrieved profiles and item attributes from the Memory Module. This process transparently reconstructs group preference aggregation through four stages: (1) Group Interests Collection, (2) Group Consensus Refinement, (3) Multi-dimensional Evaluation, and (4) Explainable Recommendation Generation, thereby moving beyond ``black-box" inference to deliver fully interpretable recommendations. Similar to the design of the Memory Module, we also implement the Reasoning Module as functions that can be invoked directly by LLMs (shown in Fig.~\ref{fig:Memory_Reasoning_Interface}). The details of each reasoning stage are as follows, with the complete prompt illustrated in Fig.~\ref{fig:prompts}.

\subsubsection{Stage 1: Group Interests Collection}
At time $t$, given a target group $g_j$, the Reasoning Module retrieves its corresponding member profiles $\{ P_{u_{1}},\ldots,P_{u_{m}}\}$ from the Memory Module. It then extracts high-frequency attributes (e.g., genre, theme) for each member $u_q$, and subsequently synthesizes these into a cohesive set of interest labels representing the collective interests of the group.

\subsubsection{Stage 2: Group Consensus Refinement}
Based on the aggregated group interests, this stage refines the group's collective preference profile through the following analytical steps:
\begin{itemize}[leftmargin=5pt]
	\item \textit{Identify Common Tags}: Compute the intersection or high-frequency union of all member tags to obtain group common tag set $\mathcal{T}_{g}$.
	\item \textit{Analyze Preference Distribution}: Examine the distribution of member preferences to identify core interests and peripheral differences.
	\item \textit{Evaluate Preference Compatibility}: Assess potential conflicts and complementarities between different member preferences.
	\item Finally, output the group consensus profile $P_g$.
\end{itemize}

\hide{
	\subsubsection{Stage 3: Multi-dimensional Evaluation}
		At this stage, we employ the LLM-as-a-judge approach~\cite{DBLP:journals/corr/abs-2411-15594} to evaluate each item $i$ in the candidate set $\mathcal{C}$ along three dimensions:
		\begin{itemize}[leftmargin=5pt]
			\item \textit{Group Common Tag Relevance Score $s_1$}: Measures the semantic similarity between the attributes of item and the group's common tag set $\mathcal{T}_g$.
			\item \textit{Member Preference Coverage Score $s_2$}: Evaluate the proportion of individual member preferences $P_{u_q}$ that are satisfied by item $i$.
			\item \textit{Recommendation Diversity Score $s_3$}: Calculate the dissimilarity between item $i$ and the group's historical interactions to enhance recommendation diversity.
		\end{itemize}
		Each item receives an evaluation vector $\textbf{s}(i) = [ s_1(i), s_2(i), s_3(i)]$.}

\subsubsection{Stage 3: Multi-dimensional Evaluation}
At this stage, to reduce judgement bias in LLM evaluation, we employ the In-Context Learning (ICL) + LLM-as-a-judge approach~\cite{DBLP:journals/corr/abs-2411-15594} to evaluate each item $i$ in the candidate set $\mathcal{C}$ along three dimensions. Specifically, we first recruit academic volunteers with a background in recommender systems to annotate 100 examples. Then, for each group $g_j$ and its candidate set $\mathcal{C}_{j}$, we randomly sample 20 out of these 100 examples, which serve as few-shot ground-truth labels to guide the LLM in scoring $s_1$, $s_2$, and $s_3$.
The detailed scoring criteria for the three dimensions are defined as follows:
\begin{itemize}[leftmargin=5pt]
	\item \textit{Group Common Tag Relevance Score $s_1$}: Measures the semantic similarity between item attributes and the group's common tag set $\mathcal{T}_g$
	(0--1: irrelevant, 2--3: partially relevant, 4--5: highly relevant).
	\item \textit{Member Preference Coverage Score $s_2$}: Evaluates the proportion of members whose preferences $P_{u_q}$ are satisfied
	(0--1: 0\% coverage, 2--3: 1--2 members, 4--5: $\ge$3 members).
	\item \textit{Recommendation Diversity Score $s_3$}: Calculates the novelty and dissimilarity between the item and the group's historical interactions
	(0--1: fully identical, 2--3: slightly novel, 4--5: highly novel).
\end{itemize}
Each item finally obtains an evaluation vector $\textbf{s}(i) = [s_1(i), s_2(i), s_3(i)]$.

\hide{	the LLM-as-a-judge paradigm to evaluate each item $i$ in the candidate set $\mathcal{C}$.
	To alleviate potential bias from vanilla LLM scoring, we introduce human annotation and In-Context Learning (ICL), and compare three implementations:
	1) \textit{Pure LLM-as-a-judge}: Directly scoring by LLM without any human annotation;
	2) \textit{LLM-as-a-judge + Human Annotation}: Scoring with reference to human-annotated ground truth;
	3) \textit{LLM-as-a-judge + In-Context Learning}: Annotating a small number of samples manually and using them as few-shot examples in the prompt to guide LLM scoring.
	Specifically, we invite 3 annotators to independently score 100 samples to obtain ground-truth labels for $s_1$, $s_2$, and $s_3$.
	These annotated samples are used as few-shot examples for in-context learning.
	Ablation results show that the third approach, \textit{LLM-as-a-judge + ICL}, achieves the best performance and is adopted as the default setting in this paper.
	The consistency between human annotations and LLM judgments yields a Cohen's Kappa coefficient of 0.81--1.00, demonstrating high reliability.
	This verifies the effectiveness of ICL in reducing LLM judgment bias, which is why we adopt this strategy throughout all experiments.}

\subsubsection{Stage 4: Explainable Recommendation Generation}
Finally, we rank candidate items based on their evaluation vector set $\{\textbf{s}(i_1), \textbf{s}(i_2)$, $\cdots \}$. Formally, we compute a composite score using a weighted score function, i.e., $F(\textbf{s}(i)) = \sum_{k=1}^{3} a_k s_k(i)$, where $a_k$ is the weight for the $k$-th evaluation dimension. The top-$k$ items with the highest scores constitute the final recommendation list $ \{ i_1, i_2, \cdots, i_k\}$. Then, we generate explanatory text $e_i$ for each recommended item $i$. Each explanation addresses three key aspects: 1) The relevance of item $i$ to the group's common tags $\mathcal{T}_g$; 2) How $i$ satisfies or balances individual preferences of specific members; 3) The contribution of $i$ in terms of recommendation diversity and novelty. This multi-stage process enhances the transparency of recommendations by providing users with a clear, step-by-step rationale for each decision.

\subsection{Model Training}
We adopt the two-stage RFT paradigm~\cite{DBLP:conf/nips/SchickDDRLHZCS23}, beginning with SFT to establish task comprehension, followed by GRPO to reinforce the agent's strategic coordination of memory and reasoning.

\subsubsection{SFT Stage}
We construct high-quality training data via human-AI collaboration, using a combination of self-refinement~\cite{DBLP:conf/nips/MadaanTGHGW0DPY23} and In-Context Learning (ICL)~\cite{DBLP:conf/nips/BrownMRSKDNSSAG20}.
First, we use DeepSeek-R1~\cite{DBLP:journals/corr/abs-2501-12948} to generate initial recommendations and explanations for a given group $g_j$ and its interaction history $\mathcal{A}_j$.
The model then self-evaluates its output and assesses recommendation difficulty, guided by 20 manually annotated few-shot examples from experts, with dynamic group-wise sampling. The evaluation produces a standardized difficulty score by examining logic consistency, explanation rationality, and overall task difficulty. Based on the critique, the model refines its outputs over two iterations. From all refined candidates, we then apply dynamic group-wise sampling, i.e., stratifying samples by difficulty scores, to select 500 representative hard examples for manual annotation, ensuring a high-quality and diverse ground truth.
Three domain experts independently annotate each sample, and majority voting determines the final label, ensuring reliability and consistency. The annotation covers two dimensions: the accuracy of the recommendation list (HR@5) and the quality of the explanation (1--5 score).
Finally, we randomly sample 10 examples from the 500 expert-calibrated ground-truth instances as in-context examples. Then, we apply the ICL approach to batch-produce the high-quality training corpus.
We represent each constructed data instance as $(q_j, \mathcal{A}_j, o_j)$,
where $q_j$ denotes the basic group information, $\mathcal{A}_j$ denotes interaction history, and $o_j$ constitutes the expert-calibrated ground-truth group recommendation.
The SFT objective is:
$\mathcal{L}_{SFT} = - \mathbb{E}_{(q_j, \mathcal{A}_j, o_j) \sim \mathcal{D}} \sum_{t=1}^{T} \log P_{\theta}( y_t \mid x )$,
where $\mathcal{D}$ is the training dataset, $x = \{ q_j, \mathcal{A}_j\}$ denotes the model input, and $y = \{o_j\}$ denotes the supervision signal.

\subsubsection{RL Stage}
\label{sec:rl}
\begin{figure}[h!]
	\centering
	\includegraphics[width=0.48\textwidth]{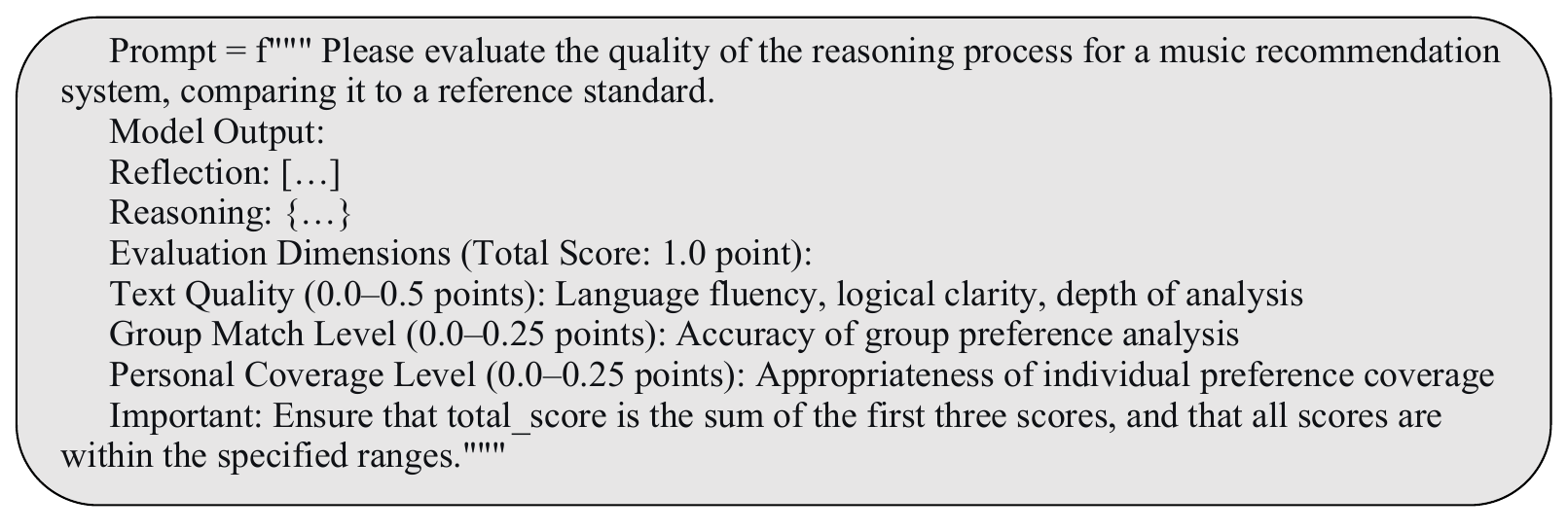}
	\caption{Reasoning reward prompt.}
	\label{fig:judge_prompt}
\end{figure}

To equip the model with the ability to autonomously invoke the Memory and Reasoning Modules for group recommendation, we further optimize its policy using the GRPO algorithm~\cite{DBLP:journals/corr/abs-2402-03300}. Formally, for each input state $x$, GRPO samples a set of actions from the current policy $\pi_{\theta}$ to generate a group of candidate answers $\{ o_1, o_2, \ldots, o_G \}$, where each $o_i \sim \pi_{\theta}( o \mid x ), i = 1,2,\ldots,G$. This sampling strategy encourages the model to explore diverse responses, preventing premature homogenization of outputs. A reward model then scores each generated answer, producing a reward sequence $\{ r_1, r_2, \ldots, r_G \}$. To guide the model toward effectively utilizing the Memory and Reasoning Modules, we design the following reward functions:

\begin{itemize}[leftmargin=10pt]
	\item \textbf{Format Reward $R_f$}: To enforce structured reasoning and output consistency, we verify the presence of required process tags: \texttt{<think>}, \texttt{<memory>}, \texttt{<reasoning>}, and \texttt{<rec>}. Each correctly included tag contributes 0.25 points, for a maximum of 1.0.
	\item \textbf{Recommendation Reward $R_{rec}$}: For each generated answer \(o_j\), we compute its NDCG@\(k\) and HitRatio@\(k\) relative to the ground-truth recommendation list. The reward is defined as: $R_{rec}^{j} = 0.5 \cdot \text{NDCG}@k + 0.5 \cdot \text{HitRatio}@k$.
	\item \textbf{Reasoning Reward $R_{rea}$}: To alleviate bias from vanilla LLM evaluation, we employ LLM-as-a-judge calibrated by human-annotated few-shot examples and In-Context Learning (ICL). The score (0–1) has three weighted dimensions: Text Generation Quality (weight 0.5), Group Preference Alignment (weight 0.25), and Personalized Preference Coverage (weight 0.25). Fig.~\ref{fig:judge_prompt} shows the evaluation prompt.
\end{itemize}
The final reward is calculated as a weighted sum: $r = w_1 R_f + w_2 R_{rec} + w_3 R_{rea}$, where $w_1, w_2, w_3$ are tunable hyperparameters that balance the contributions of each type of reward.
Finally, we maximize the GRPO objective: $\mathcal{J}(\theta)= \mathbb{E}_{x\sim\mathcal{D}, \{ o_i \}_{i=1}^{G} \sim \pi_{\text{old}}( \cdot \mid x)} 
\frac{1}{N} \sum_{i=1}^{G} \\
\min\Biggl( \frac{\pi_{\theta}( o \mid x )}{\pi_{\text{old}}( o \mid x )}A_i,
\text{clip}\Bigl(\frac{\pi_{\theta}( o \mid x )}{\pi_{\text{old}}( o \mid x )}, 1-\varepsilon, 1+\varepsilon\Bigr)A_i \Biggr)$, where $A_i = \frac{r_i - \text{mean}(r)}{\text{std}(r)}$ is the advantage function.

	\hide{\section{Theoretical Analysis}
\label{sec:theory}
The predominant method for training LLM agents employs the SFT+GRPO paradigm~\cite{DBLP:journals/corr/abs-2501-12948}. Prior work~\cite{DBLP:conf/nips/RafailovSMMEF23,DBLP:journals/tmlr/ZhangGYYZTZLXLZCZFWHVLW26} shows that SFT+GRPO outperforms other common paradigms, such as SFT+DPO, as well as individual methods like SFT, DPO, or GRPO. However, the theoretical basis for why SFT+GRPO is superior to SFT+DPO has not been explored. For this reason, we have provided a theoretical proof.

\noindent \textbf{Lemma 1}: The DPO loss function $\mathcal{L}_{\text{DPO}}(\theta) $ is defined as:
$-\mathbb{E}_{(x, y_w, y_l) \sim D} \left[ \log \sigma\left( \beta \log \frac{\pi_\theta(y_w \mid x)}{\pi_{\text{SFT}}(y_w \mid x)} - \beta \log \frac{\pi_\theta(y_l \mid x)}{\pi_{\text{SFT}}(y_l \mid x)} \right) \right]$,
where \(y_w\) denotes a preferred response, \(y_l\) denotes a dispreferred response, \(\pi_{\text{SFT}}\) denotes the model resulting from SFT, \(\pi_\theta\) is the model to be optimized, and \(\beta\) is a hyperparameter.

\subsection{Generalization Error Bound Analysis}
\noindent \textbf{Theorem 1}: Let $\text{Bound}_{\text{DPO}}$ and $\text{Bound}_{\text{GRPO}}$ denote the generalization error bounds for DPO and GRPO, respectively. When the GRPO sample count satisfies $K \geq 4$, the inequality $\text{Bound}_{\text{DPO}} > \text{Bound}_{\text{GRPO}}$ holds tightly. Therefore, when combining SFT for downstream task training, the performance of SFT+GRPO will be superior to that of SFT+DPO.

\noindent \textbf{Proof}:
\noindent\textbf{Step 1: Derivation of $\text{Bound}_{\text{DPO}}$.}
Given input \(x\), DPO~\cite{DBLP:conf/nips/RafailovSMMEF23} samples \(N_{\mathrm{pair}}\) independent pairs \((y_w^{(i)}, y_l^{(i)})\). Its empirical error rate \(\widehat{R}_{\mathrm{DPO}}(r)\) is:
$
\widehat{R}_{\mathrm{DPO}}(r) = \frac{1}{N_{\mathrm{pair}}}\sum_{i=1}^{N_{\mathrm{pair}}} \mathbb{I}\left[ r(y_w^{(i)}) \leq r(y_l^{(i)}) \right]
$,
where \( r \in \mathcal{F} \) is the reward function, \( |\mathcal{F}| < \infty \), and \( r(y_w^{(i)}) \) and \( r(y_l^{(i)}) \) represent the scores assigned by DPO to the preferences \( y_w^{(i)} \) and \( y_l^{(i)} \), respectively. For a fixed reward function \( r \), \( \widehat{R}_{\mathrm{DPO}}(r) \) is the mean of \( N_{\mathrm{pair}} \) independent and identically distributed random variables, taking values in \([0,1]\). By using Hoeffding's inequality~\cite{HoeffdingW963/6}, we have $
\mathbb{P}\left( |\widehat{R}_{\mathrm{DPO}}(r) - R_{\mathrm{const}}(r)| \geq \varepsilon \right) \leq 2 \exp\left( -2 N_{\mathrm{pair}} \varepsilon^2 \right),
$
where \( R_{\mathrm{const}}(r) \) denotes the true error rate at the constraint level:
$
R_{\mathrm{const}}(r) = \mathbb{E}_{y_w \sim p^+, y_l \sim p^-} \left[ r(y_w) \leq r(y_l) \right],
$
with \( p^+ \) and \( p^- \) representing the true positive and negative distributions, respectively.

\noindent Applying the union bound over all \( r \in \mathcal{F} \):
$
\mathbb{P}( \exists r \in \mathcal{F}: |\hat{R}_{\mathrm{DPO}}(r) - R_{\mathrm{const}}(r)| \geq \varepsilon ) \leq 2 |\mathcal{F}| \exp( -2 N_{\mathrm{pair}} \varepsilon^2 ).
$

\noindent Setting \( 2 |\mathcal{F}| \exp\left( -2 N_{\mathrm{pair}} \varepsilon^2 \right) = \delta \), we obtain:
$
\varepsilon = \sqrt{\frac{\ln(2 |\mathcal{F}| / \delta)}{2 N_{\mathrm{pair}}}}.
$

\noindent Thus, with probability at least \( 1 - \delta \), for all \( r \in \mathcal{F} \):
$
R_{\mathrm{const}}(r) \leq \widehat{R}_{\mathrm{DPO}}(r) + \sqrt{\frac{\ln(2 |\mathcal{F}| / \delta)}{2 N_{\mathrm{pair}}}}.
$
That is:
$
\sup_{r \in \mathcal{F}} |R_{\mathrm{const}}(r) - \widehat{R}_{\mathrm{DPO}}(r)| \leq \sqrt{\frac{\ln(2 |\mathcal{F}| / \delta)}{2 N_{\mathrm{pair}}}}.
$
Here, \( \sup \) denotes the supremum, and \( |R_{\mathrm{const}}(r) - \widehat{R}_{\mathrm{DPO}}(r)| \) represents the generalization error. Therefore, the generalization error upper bound for DPO is: $
\mathrm{Bound}_{\mathrm{DPO}} = \sqrt{\frac{\ln(2 |\mathcal{F}| / \delta)}{2 N_{\mathrm{pair}}}}.
$

\noindent\textbf{Step 2: Derivation of $\text{Bound}_{\text{GRPO}}$.}
GRPO samples \( N_{\mathrm{group}} \) groups, each containing \( Y^+ = \{ y_1^+, \dots, y_m^+ \} \) and \( Y^- = \{ y_1^-, \dots, y_n^- \} \), yielding \( m n \) constraints \( (y^+, y^-) \) per group.

\noindent For a fair comparison at the constraint level, the empirical error rate for GRPO is defined as the average 0-1 loss over all constraints across all groups:
\[
\widehat{R}_{\mathrm{GRPO}}(r) = \frac{1}{N_{\mathrm{group}} \cdot m n} \sum_{t=1}^{N_{\mathrm{group}}} \sum_{i=1}^{m} \sum_{j=1}^{n} \mathbb{I}\left[ r(y_{t,i}^+) \leq r(y_{t,j}^-) \right].
\]

\noindent For a fixed reward function \( r \), \( \widehat{R}_{\mathrm{GRPO}}(r) \) is the mean of \( N_{\mathrm{group}} \cdot m n \) independent and identically distributed random variables, taking values in \([0,1]\). By Hoeffding's inequality:
$
\mathbb{P}\left( |\widehat{R}_{\mathrm{GRPO}}(r) - R_{\mathrm{const}}(r)| \geq \varepsilon \right) \leq 2 \exp\left( -2 (N_{\mathrm{group}} \cdot m n) \varepsilon^2 \right).
$
Applying the union bound over all \( r \in \mathcal{F} \):
$
\mathbb{P}\left( \exists r \in \mathcal{F}: |\widehat{R}_{\mathrm{GRPO}}(r) - R_{\mathrm{const}}(r)| \geq \varepsilon \right) \leq 2 |\mathcal{F}| \exp\left( -2 N_{\mathrm{group}} \cdot m n \varepsilon^2 \right).
$
Setting \( 2 |\mathcal{F}| \exp\left( -2 N_{\mathrm{group}} \cdot m n \varepsilon^2 \right) = \delta \), we obtain:
$
\varepsilon = \sqrt{\frac{\ln(2 |\mathcal{F}| / \delta)}{2 N_{\mathrm{group}} m n}}.
$
Thus, with probability at least \( 1 - \delta \), for all \( r \in \mathcal{F} \):
$
R_{\mathrm{const}}(r) \leq \widehat{R}_{\mathrm{GRPO}}(r) + \sqrt{\frac{\ln(2 |\mathcal{F}| / \delta)}{2 N_{\mathrm{group}} m n}}.
$
That is:
$
\sup_{r \in \mathcal{F}} |R_{\mathrm{const}}(r) - \widehat{R}_{\mathrm{GRPO}}(r)| \leq \sqrt{\frac{\ln(2 |\mathcal{F}| / \delta)}{2 N_{\mathrm{group}} m n}}.
$
Therefore, the generalization error upper bound for GRPO is:
$
\mathrm{Bound}_{\mathrm{GRPO}} = \sqrt{\frac{\ln(2 |\mathcal{F}| / \delta)}{2 N_{\mathrm{group}} m n}}.
$

\noindent\textbf{Step 3: Fair Comparison.}
Under the same total output sample budget, DPO produces two outputs per pair \((y_w, y_l)\), resulting in \( 2 N_{\mathrm{pair}} \) total outputs, while GRPO produces \( N_{\mathrm{group}} \cdot (m + n) = N_{\mathrm{group}} \cdot K \) outputs, where \( m + n = K \). Equating the total outputs:
\[
2 N_{\mathrm{pair}} = N_{\mathrm{group}} \cdot K \quad \Rightarrow \quad N_{\mathrm{group}} = \frac{2 N_{\mathrm{pair}}}{K}.
\]

\noindent Substituting into \( \mathrm{Bound}_{\mathrm{GRPO}} \)
$
= \sqrt{\frac{\ln(2 |\mathcal{F}| / \delta)}{2 \cdot \frac{2 N_{\mathrm{pair}}}{K} \cdot m n}} = \sqrt{\frac{\ln(2 |\mathcal{F}| / \delta)}{4 N_{\mathrm{pair}} \cdot \frac{m n}{K}}}.
$
Thus, the ratio of the bounds is:
$
\frac{\mathrm{Bound}_{\mathrm{GRPO}}}{\mathrm{Bound}_{\mathrm{DPO}}} = \sqrt{\frac{2 N_{\mathrm{pair}}}{4 N_{\mathrm{pair}} \cdot \frac{m n}{K}}} = \sqrt{\frac{1}{2} \cdot \frac{K}{m n}}.
$

\noindent We now compare \( K \) and \( 2 m n \), i.e., compare \( m + n \) and \( 2 m n \). For simplicity, set \( m = n \). Then:
$
K = 2m, \quad \frac{K}{m n} = \frac{2m}{m^2} = \frac{2}{m}, \quad \frac{1}{2} \cdot \frac{K}{m n} = \frac{1}{m}.
$
Therefore:
$
\frac{\mathrm{Bound}_{\mathrm{GRPO}}}{\mathrm{Bound}_{\mathrm{DPO}}} = \sqrt{\frac{1}{m}}.
$
Since \( m \geq 1 \), we have \( \frac{\mathrm{Bound}_{\mathrm{GRPO}}}{\mathrm{Bound}_{\mathrm{DPO}}} \leq 1 \). Consequently, when \( m = n \geq 2 \), i.e., when the GRPO rollout \( K \geq 4 \), $\mathrm{Bound}_{\mathrm{GRPO}} $ is strictly smaller than $\mathrm{Bound}_{\mathrm{DPO}}$. 

\textit{Remarks.} Our analysis assumes equal sample budgets and finite \(\mathcal{F}\) (standard in \cite{DBLP:conf/nips/RafailovSMMEF23,DBLP:journals/corr/abs-2402-03300}); empirical results (Fig.~\ref{fig:rollout}) validate the predicted advantage.
}
	\section{Experiments}
We conduct comprehensive experiments to answer the following Research Questions (RQs):
\begin{itemize}[leftmargin=10pt]
	\item \textbf{RQ1}: Is AGR superior to the baseline method in terms of recommendation accuracy and interpretability?
	\item \textbf{RQ2}: Does AGR effectively capture the temporal evolution of group preferences across multiple prediction steps, and how does it compare to baselines?
	\item \textbf{RQ3}: How do the Memory Module's modeling of evolving profiles and the Reasoning Module's multi-stage process contribute to a more accurate and interpretable recommendation system? 
	\item \textbf{RQ4}: What is the necessity of the SFT and RL phases within the RFT training process? 
	\item \textbf{RQ5}: How does reward design influence the model?
	\item \textbf{RQ6}: What is the efficiency of the Memory Module's core operations (insertion, updating, forgetting, summarization, and retrieval), and what is the efficiency of each stage within the Reasoning Module? 
	\item \textbf{RQ7}: How do hyperparameters impact overall performance?
\end{itemize}

\subsection{Experimental Setup}
\subsubsection{Datasets}
We construct our dataset based on two public group recommendation benchmarks, LastFM~\cite{DBLP:journals/is/KimE15} and Douban~\cite{DBLP:journals/ijdsa/ZhengLSZLW17}, where LastFM recommends music artists, Douban provides movie recommendations. In both datasets, the items for group interaction are chronologically ordered.
Following the mainstream method~\cite{DBLP:conf/icde/YinW0LYZ19,DBLP:conf/sigir/Cao0MAYH18,DBLP:conf/aaai/Ye0WCZ025} of group partitioning in group recommendation, we form groups based on user interactions and social relationships, under the following rules. For LastFM, two or more users are formed into a group if they share a social friendship connection and have at least five common artists in their interaction history. For Douban, two or more users are formed into a group if they have collectively watched at least five identical movies. The group size for both datasets is controlled between 3 and 8 users, with an average size of 4.2 users, to ensure group diversity and reliable modeling of temporal dependencies. For dataset splitting, we adopt the widely-used Leave-One-Out protocol~\cite{DBLP:conf/sigir/Cao0MAYH18}: for each user/group interaction sequence, all interactions except the last one are used as training history, and the last interaction is used as the test target to predict the next item. Dataset statistics are summarized in Table~\ref{tab:datastats}.

\begin{table}[h!]
	\centering
	\caption{Dataset statistics. U-Inter. means total user interactions; G-Inter. means total group interactions.}
	\label{tab:datastats}
	\begin{tabular}{lcccccc}
		\toprule
		Dataset & Users & Groups & U-Inter. & G-Inter. & Tags \\
		\midrule
		LastFM & 1,892 & 5,529 & 92,834 & 31,665 & 11,946 \\
		Douban & 13,545 & 6,141 & 199,813 & 51,472 & 199,813 \\
		\bottomrule
	\end{tabular}
\end{table}

\subsubsection{Baselines}
Our evaluation includes nine categories of comparative methods:
(1) Deep learning-based methods (AGREE~\cite{DBLP:conf/sigir/Cao0MAYH18}, DisREC~\cite{DBLP:conf/aaai/Ye0WCZ025});
(2) GNN-based group-aware recommendation method: GGRM~\cite{DBLP:journals/tois/Tian0SJZ22};
(3) Time-aware sequential recommendation methods: SASRec~\cite{DBLP:conf/icdm/KangM18}, GRU4Rec~\cite{DBLP:conf/aciids/DobrovolnySK21};
(4) LLM-based temporal recommendation method: LOHRec
~\cite{DBLP:conf/emnlp/XieWJYM25};
(5) Memory and reasoning-augmented LLMs: MemoCRS~\cite{DBLP:conf/cikm/XiLL0T0024}, MR.Rec
~\cite{DBLP:journals/corr/abs-2510-14629} ,which employ predefined strategies to manage memory through insert, delete, update, query, forget, and summarize operations in response to evolving user behaviors;
(6) AGR-API: Calls the LLM API (DeepSeek-R1~\cite{DBLP:journals/corr/abs-2501-12948}) with user histories and item descriptions as text prompts, using structured templates to trigger external tools (e.g., memory retrieval) via function calling. Notably, it keeps LLM interpretability and basic tool invocation capabilities, but it lacks the dynamic memory and multi-stage reasoning of the full AGR framework.
(7) LLM-based methods ~\cite{DBLP:conf/um/LubosFGLHWF25} that use vanilla LLMs (DeepSeek-R1~\cite{DBLP:journals/corr/abs-2501-12948}, LLaMA3~\cite{DBLP:journals/corr/abs-2407-21783});
(8) CoT+ICL~\cite{DBLP:conf/acl/KothapalliFS25} that uses Chain-of-Thought prompts to infer group preferences step-by-step, then provides top-$k$ recommendations and explanations;
(9) Our proposed AGR framework and its variants: AGR-M (memory only), AGR-R (reasoning only), AGR-SFT (SFT only), AGR-GRPO (GRPO only).

\subsubsection{Evaluation Metrics}
The selected evaluation metrics fall into three categories: recommendation performance metrics (HR@$k$, NDCG@$k$, MRR), explainability metrics (BLEU, ROUGE*, BERTScore, Hallucination Rate), and human evaluation scores (CPA, PDA, RQS, EQ), where ROUGE* denotes the average score of ROUGE-1, ROUGE-2, and ROUGE-L.
CPA, PDA, and RQS primarily assess recommendation performance, while EQ is designed to evaluate explanation quality. More concretely,
CPA measures how accurately the model summarizes group common preferences, calculated as $\text{CPA} = \frac{1}{N}\sum S_{\text{cpa}}$; PDA evaluates the model's ability to identify and explain preference disparities among members, defined as $\text{PDA} = \frac{1}{N}\sum S_{\text{pda}}$; RQS comprehensively assesses recommendation lists across four dimensions (group matching, member coverage, diversity, novelty), formulated as $\text{RQS} = \sum_{k=1}^{4} \omega_k \cdot \text{dim}_k$ with weights typically set to 0.25; and EQ evaluates explanation quality in terms of transparency ($S_T$), comprehensibility ($S_C$), and persuasiveness ($S_P$), i.e.,  $\text{EQ} = 0.4 \cdot S_T + 0.3 \cdot S_C + 0.3 \cdot S_P$. Notably, $S_{\text{cpa}}$ and $S_{\text{pda}}$ are human evaluation scores on a 10-point scale, further divided into four levels: Poor (0–2), Fair (3–5), Good (6–8), and Excellent (9–10). Human evaluations were conducted by three professional domain experts in a single-blind manner. Each sample was independently evaluated twice by each expert. Inter-rater agreement among the three experts was measured using Fleiss' Kappa~\cite{fleiss1971measuring}, which is designed for agreement assessment with multiple raters ($\kappa = 0.74$), confirming the reliability of the evaluation.

\subsubsection{Implementation Details}
Due to the limited input length of LLMs, all models are evaluated under a constrained ranking setting to ensure a fair comparison. Specifically, we pre-filter the item pool by performing Ranking Retrieval (as described in Section~\ref{sec:memory}) to obtain the top-$K$ ($K$=200) most semantically relevant candidate set $\mathcal{C}$. For fair comparison, all baselines are evaluated using this identical candidate set $\mathcal{C}$ to maintain consistency under this protocol. We use LLaMA3
-8B~\cite{DBLP:journals/corr/abs-2407-21783} as the base LLM, with DeepSeek-R1-Distill-Qwen-7B \cite{DBLP:journals/corr/abs-2501-12948} used to assess generalization. Fine-Tuning via LoRA was conducted on dual NVIDIA RTX 4090 GPUs. Key configurations include a learning rate of 5e-6, temperature of 0.7, training/inference batch size of 1 and 4, GRPO $\beta$ of 0.001, a rollout of 4, a 2,048-token sequence limit, and uniform reward weights $w_1, w_2, w_3=(1,1,1)$.

\begin{figure}[h]
	\centering
	\subfigure[BERTScore]{\label{subfig:deepseek1.pdf}
		\includegraphics[width=0.22\textwidth]{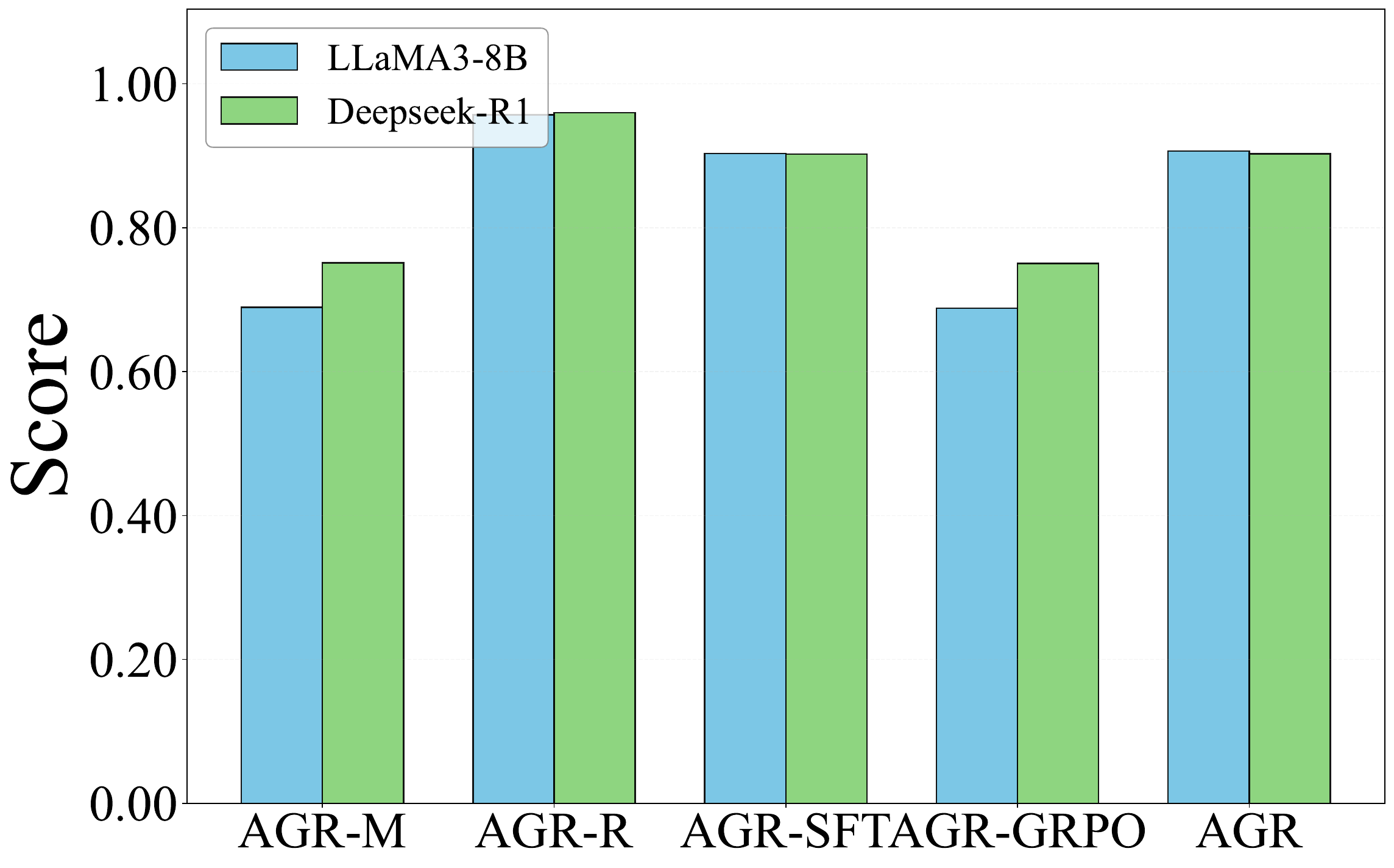}}
	\subfigure[HR@10]{\label{subfig:deepseek2.pdf}
		\includegraphics[width=0.22\textwidth]{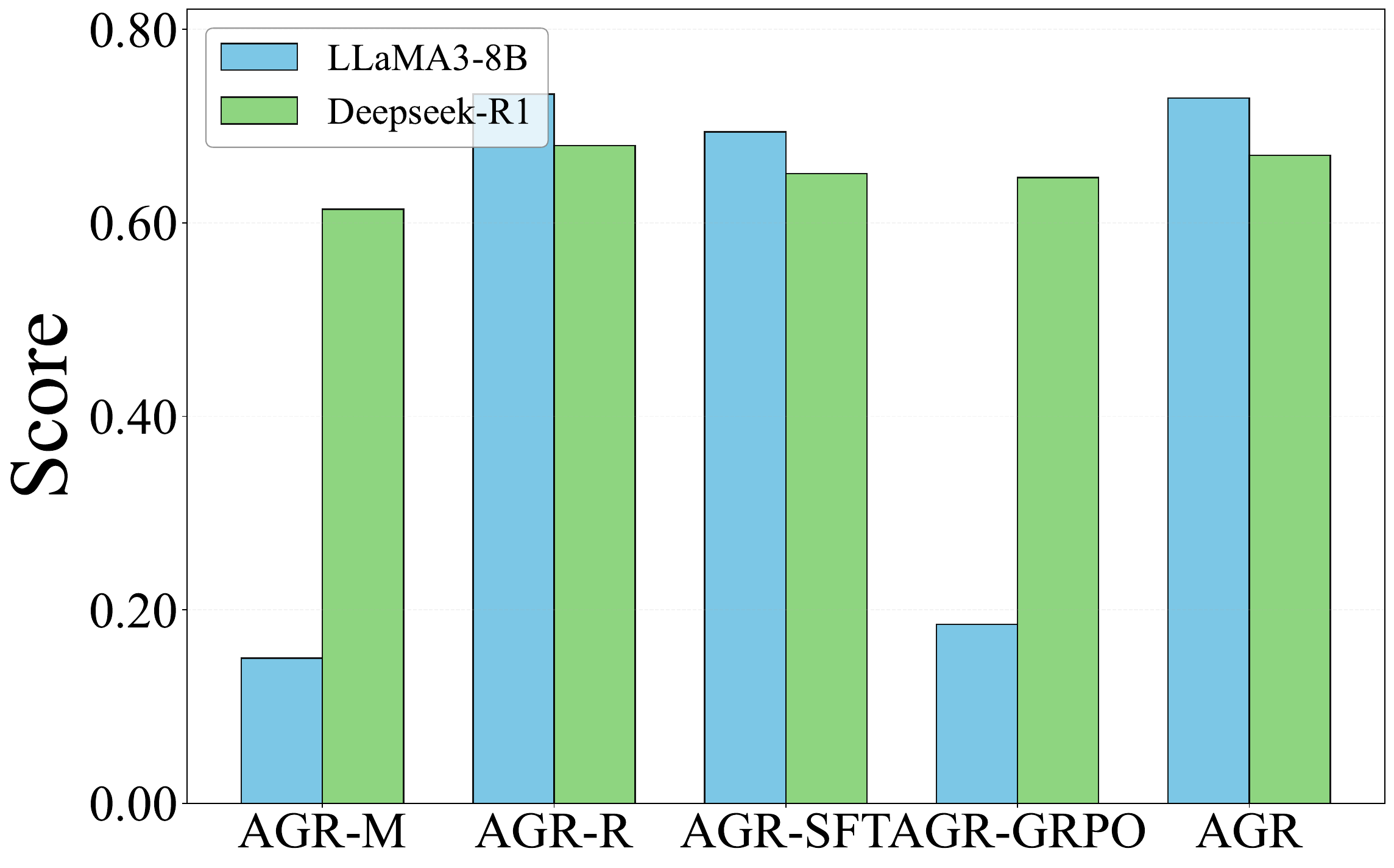}}	
	\caption{Comparison of different backbone LLMs, $p < 0.01$.}
	\label{fig:deepseek}
	\vspace{-6pt}
\end{figure}

\subsection{Overall Performance (RQ1 \& RQ2)}
\begin{figure}[h]
	\centering
	\subfigure[BLEU]{\label{subfig:lastfm_bleu}
		\includegraphics[width=0.45\linewidth]{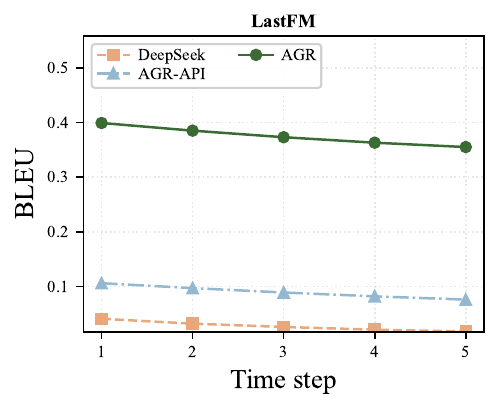}}
	\hfill
	\subfigure[BLEU]{\label{subfig:douban_bleu}
	\includegraphics[width=0.45\linewidth]{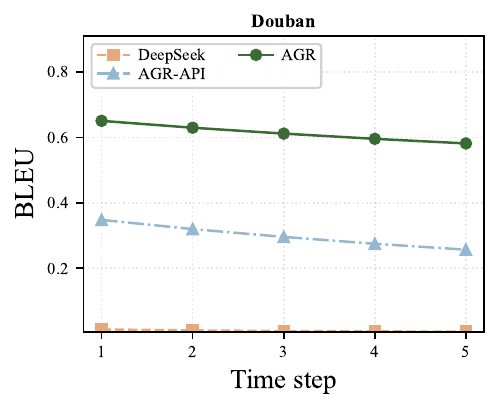}}
	\subfigure[HR@5]{\label{subfig:lastfm_hr5}
		\includegraphics[width=0.45\linewidth]{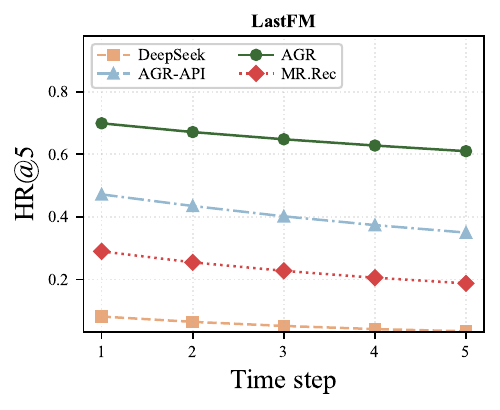}}
	\hfill
	\subfigure[HR@5]{\label{subfig:douban_hr5}
		\includegraphics[width=0.45\linewidth]{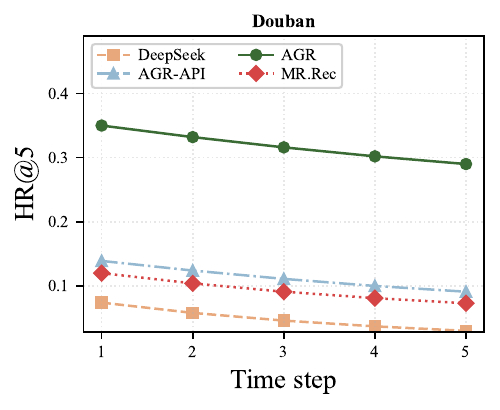}}
	\subfigure[CPA]{\label{subfig:lastfm_CPA}
	\includegraphics[width=0.45\linewidth]{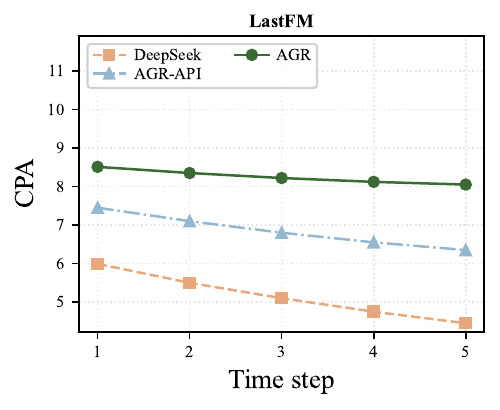}}
	\hfill
	\subfigure[CPA]{\label{subfig:douban_CPA}
	\includegraphics[width=0.45\linewidth]{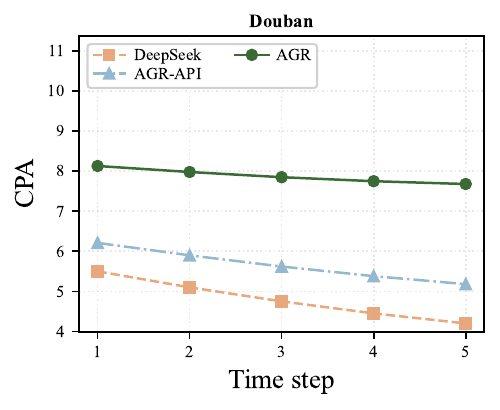}}
	\caption{Temporal evaluation on Douban and LastFM. The x-axis $t$ denotes the number of held-out future interactions (leave-last-$t$-out).}
	\label{fig:rq8}
\end{figure}

The overall performance on the LastFM and Douban datasets across all methods is shown in Tables~\ref{tb:overall_lastfm_t}–\ref{tb:overall_douban_t}. All performance improvements are statistically significant with $p < 0.001$ using the Mann-Whitney U test~\cite{Mann1947OnAT}.
Our findings are as follows:
1) \sRC achieves the best performance, which implies it can effectively model the dynamic evolution of group/user preferences and can explicitly explain the group decision-making process.
2) AGR greatly outperforms AGR-API on all metrics, likely because the underlying LLM is a general-purpose model without specialized knowledge of group recommendation.
3) Deep learning-based models achieve competitive recommendation scores, yet their lack of explainability reduces transparency compared to LLM-based approaches.
4)Memory-based methods are inconsistent. MR.Rec performs poorly because it relies solely on retrieval without fine-grained maintenance (e.g., removing irrelevant interactions or updating profiles). MemoCRS supports profile evolution through Create, Read, Update, and Delete operations,but uses heuristic maintenance and lacks autonomous updates, making it underperform compared to AGR.
5) LLMs achieve lower recommendation scores because they are not tailored for group recommendation. Even with advanced prompting (LLaMA+CoT+ICL), they still underperform deep learning-based methods, mainly because they are not fine-tuned for this specific problem.
6) Since \sRC beats LLaMA+CoT+ICL, this shows that RFT effectively improves performance. Additionally, we evaluate the framework's sensitivity to backbone models using LLaMA3-8B and DeepSeek-R1-Distill-Qwen-7B on the LastFM dataset. Results (Fig.~\ref{fig:deepseek}) show that DeepSeek yields the best performance, indicating a positive correlation between backbone model strength and AGR effectiveness.
7) Furthermore, to assess whether AGR captures the temporal evolution of group preferences, we construct five prediction tasks for each group using a \textit{leave-last-$t$-out} strategy. Specifically, for $t \in \{ 1, 2, 3, 4, 5\}$, we remove the most recent $t$ interactions of a group and require the model to predict the \textbf{earliest among the removed items} (i.e., the $t-1$ most recent interaction), using only interactions before the earliest held-out item as input. We compare AGR, AGR-API, MR.Rec (where applicable) and vanilla DeepSeek on LastFM and Douban evaluated with HR@5, BLEU and CPA. As shown in Fig.~\ref{fig:rq8}, AGR outperforms both baselines across all five time steps, demonstrating its ability to capture evolving group preferences and maintain superior multi-step prediction performance.

\begin{table*}[t]
	\centering
	\caption{Overall performances on LastFM, the term "ROUGE*" refers to the average score of ROUGE-1/2/L.}
	\label{tb:overall_lastfm_t}
	\small
	\begin{tabular}{l c c c c c c c c c c c}
		\toprule
		Model & HR@5 & NDCG@5 & MRR & BLEU & ROUGE* & BERTScore & CPA & PDA & RQS & EQ & Hallucinations \\
		\midrule
		AGREE & 0.3474 & 0.2002& 0.1892 & - & - & - & - & - & - & - & - \\
		DisREC & 0.3227 & 0.2227 & 0.2153 & - & - & - & - & - & - & - & - \\
		\midrule
		GGRM & 0.1177 & 0.0792 & 0.0665 & - & - & - & - & - & - & - & - \\
		\midrule
		SASRec & 0.4336 & 0.3486 & 0.3202 & - & - & - & - & - & - & - & - \\
		GRU4Rec & 0.1820 & 0.1515 & 0.1413 & - & - & - & - & - & - & - & - \\
		\midrule
		LOHRec & 0.4736 & 0.3986 & 0.3162 & - & - & - & - & - & - & - & - \\
		\midrule
		MR.Rec & 0.2400 & 0.1362 & 0.1579 & - & - & - & - & - & - & - & - \\
		MemoCRS & 0.2633 & 0.1247 & 0.1617 & 0.0419 & 0.5596 & 0.4267 & 6.75 & 5.17 & 7.28 & 1.65 & 0.1045 \\
		\midrule
		LLaMA & 0.1670 & 0.1000 & 0.0938 & 0.0198 & 0.1152 & 0.6968 & 6.14 & 3.40 & 4.40 & 5.43 & 0.8772 \\
		DeepSeek & 0.0820 & 0.0488 & 0.0462 & 0.0411 & 0.0412 & 0.7566 & 5.99 & 4.24 & 4.49 & 5.15 & 0.2223 \\
		\midrule
		AGR-API & 0.4720 & 0.2785 & 0.2040 & 0.1063 & 0.5443 & 0.8850 & 7.45 & 4.73 & 7.42 & 6.47 & 0.1360 \\
		\midrule
		LLaMA+CoT+ICL & 0.4780 & 0.3054 & 0.2849 & 0.3194 & 0.5211 & 0.9571 & 8.55 & 8.26 & 7.39 & 8.18 & \textbf{0.0548} \\
		\midrule
		LLaMA+SFT+DPO & 0.4870 & 0.3198 & 0.2989 & 0.3210 & 0.5205 & 0.9575 & 8.59 & 8.02 & 7.30 & 8.13 & 0.0584 \\
		LLaMA+GRPO & 0.1710 & 0.1000 & 0.0918 & 0.0200 & 0.1154 & 0.6955 & 5.55 & 2.98 & 3.93 & 4.84 & 0.8743 \\
		\midrule
		AGR-M & 0.1020 & 0.0600 & 0.0529 & 0.0206 & 0.1090 & 0.6893 & 4.35 & 2.47 & 3.12 & 4.18 & 0.8785 \\
		AGR-R & 0.4900 & 0.3252 & 0.3055 & 0.3230 & 0.5256 & 0.9570 & \textbf{8.60} & 8.15 & 7.30 & 8.14 & 0.4669 \\
		AGR-SFT & 0.6760 & 0.5345 & 0.4898 & \textbf{0.3999} & \textbf{0.6215} & 0.9631 & 8.52 & 7.54 & 7.13 & 8.00 & 0.0603 \\
		AGR-GRPO & 0.1250 & 0.0750 & 0.0665 & 0.0198 & 0.1101 & 0.6883 & 4.09 & 2.43 & 2.96 & 4.06 & 0.8754 \\
		\textbf{AGR} & \textbf{0.6990} & \textbf{0.5634} & \textbf{0.5227} & 0.3990 & 0.6206 & \textbf{0.9672} & 8.51 & \textbf{8.30} & \textbf{7.50} & \textbf{8.50} & 0.0645 \\
		\bottomrule
	\end{tabular}
\end{table*}

\begin{table*}[t]
	\centering
	\caption{Overall performances on Douban, the term "ROUGE*" refers to the average score of ROUGE-1/2/L.}
	\label{tb:overall_douban_t}
	\small
	\begin{tabular}{l c c c c c c c c c c c}
		\toprule
		Model & HR@5 & NDCG@5 & MRR & BLEU & ROUGE* & BERTScore & CPA & PDA & RQS & EQ & Hallucinations \\
		\midrule
		AGREE & 0.0785 & 0.0508 & 0.0704 & - & - & - & - & - & - & - & - \\
		DisREC & 0.1582 & 0.1484 & 0.1616 & - & - & - & - & - & - & - & - \\
		\midrule
		GGRM & 0.1379 & 0.0833 & 0.0711 & - & - & - & - & - & - & - & - \\
		\midrule
		SASRec & 0.1485 & 0.1124 & 0.1086 & - & - & - & - & - & - & - & - \\
		GRU4Rec & 0.0931 & 0.0676 & 0.0591 & - & - & - & - & - & - & - & - \\
		\midrule
		LOHRec & 0.2092 & 0.1535 & 0.1279 & - & - & - & - & - & - & - & - \\
		\midrule
		MR.Rec & 0.0600 & 0.0355 & 0.0370 & - & - & - & - & - & - & - & - \\
		MemoCRS  & 0.0630 & 0.0293 & 0.0568 & 0.0190 & 0.3790 & 0.5101 & 4.84 & 2.76 & 5.91 & 1.25 & 0.1010 \\
		\midrule
		LLaMA & 0.0170 & 0.0113 & 0.0165 & 0.0134 & 0.1082 & 0.6606 & 4.07 & 2.26 & 3.31 & 4.21 & 0.8112 \\
		DeepSeek & 0.0740 & 0.0416 & 0.0469 & 0.0130 & 0.0697 & 0.7515 & 5.50 & 3.96 & 4.39 & 5.12 & 0.4667 \\
		\midrule
		AGR-API & 0.1390 & 0.1046 & 0.0675 & 0.3480 & 0.8040 & 0.9680 & 6.21 & 6.99 & 6.45 & 5.57 & 0.7820 \\
		\midrule
		LLaMA+CoT+ICL & 0.1780 & 0.1238 & 0.1304 & 0.4031 & 0.5962 & 0.9717 & 8.10 & 7.60 & 4.59 & 6.70 & 0.0338 \\
		\midrule
		LLaMA+SFT+DPO & 0.1500 & 0.1040 & 0.1116 & 0.4229 & 0.6104 & 0.9756 & 8.01 & 7.22 & 4.41 & 6.55 & 0.0291 \\
		LLaMA+GRPO & 0.0600 & 0.0456 & 0.0459 & 0.0139 & 0.1125 & 0.6593 & 4.29 & 2.34 & 3.52 & 4.39 & 0.7881 \\
		\midrule
		AGR-M & 0.0280 & 0.0175 & 0.0206 & 0.0136 & 0.1059 & 0.6621 & 3.89 & 2.16 & 3.25 & 4.17 & 0.8069 \\
		AGR-R & 0.1900 & 0.1273 & 0.1304 & 0.4161 & 0.6053 & 0.9711 & 8.02 & 7.55 & 4.30 & 6.58 & 0.0342 \\
		AGR-SFT & 0.3200 & 0.1685 & 0.1508 & 0.5916 & 0.7913 & 0.9756 & 8.12 & 7.62 & 5.53 & 7.06 & \textbf{0.0291} \\
		AGR-GRPO & 0.0400 & 0.0225 & 0.0218 & 0.0140 & 0.1150 & 0.6613 & 3.83 & 2.08 & 3.06 & 3.87 & 0.8523 \\
		\textbf{AGR} & \textbf{0.3500} & \textbf{0.2079} & \textbf{0.1932} & \textbf{0.6511} & \textbf{0.8354} & \textbf{0.9765} & \textbf{8.13} & \textbf{7.67} & \textbf{5.91} & \textbf{7.31} & 0.0340 \\
		\bottomrule
	\end{tabular}
\end{table*}

\begin{table*}[t]
	\centering
	\caption{Prompt Ablation Study: BERTScore and ROUGE* Consistency Under Different Prompt Variants on LastFM.}
	\label{tb:prompt_ablation}
	\begin{tabular}{lccc}
		\toprule
		\textbf{Prompt Variant} & \textbf{Description} & \textbf{BERTScore} & \textbf{ROUGE*} \\
		\midrule
		Original & Full reasoning prompt as shown in Fig. 4 & 0.967 & 0.852 \\
		Synonym Replacement & Replaced ``Consensus'' with ``Agreement'', ``Refinement'' with ``Adjustment'' & 0.943 & 0.831 \\
		Dimension Removal & Removed explicit scoring criteria for \(s_3\) (Diversity) & 0.926 & 0.817 \\
		\bottomrule
	\end{tabular}
\end{table*}

\subsection{Ablation Studies (RQ3-RQ5)}
\begin{figure}[h]
	\centering
	\subfigure[Fleiss' Kappa]{\label{subfig:kappa}
		\includegraphics[width=0.21\textwidth]{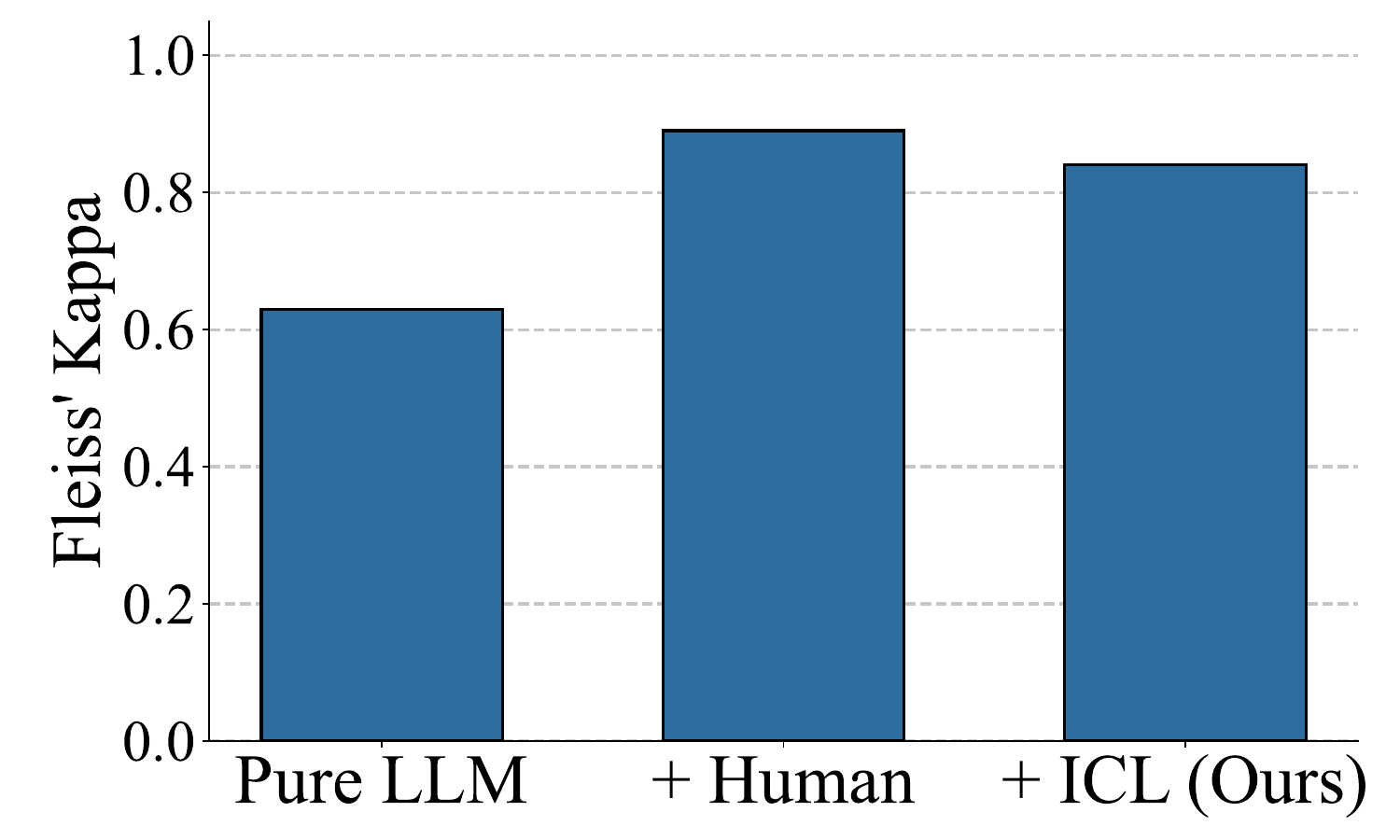}}	
	\subfigure[HR@5]{\label{subfig:hr5}
		\includegraphics[width=0.21\textwidth]{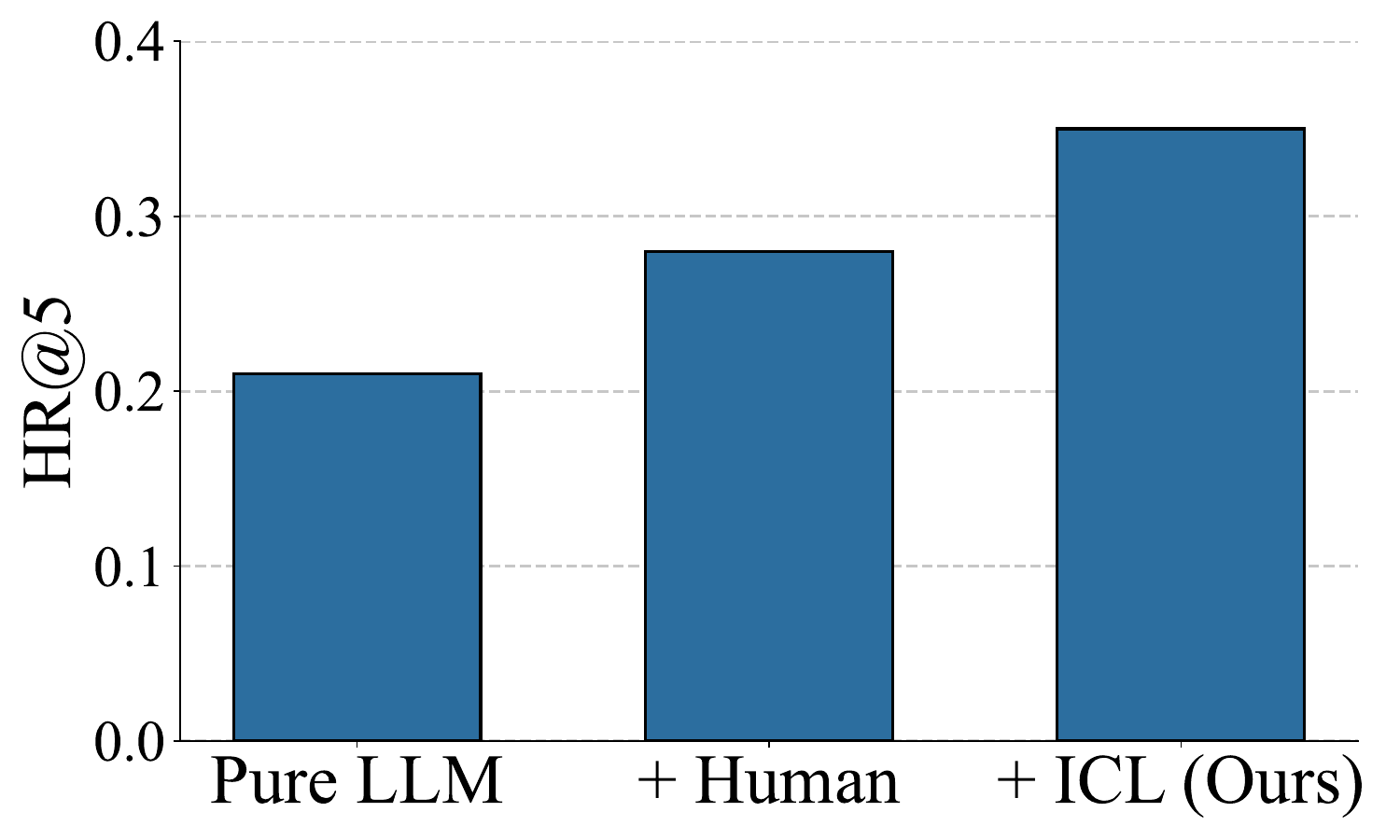}}	
	\caption{Ablation study of judgement bias analysis, $p < 0.01$.}
	\label{fig:ablation_lms}
	\vspace{-7pt}
\end{figure}

\begin{figure}[h]
	\centering
	\subfigure[ROUGE*]{\label{subfig:reward2.pdf}
		\includegraphics[width=0.21\textwidth]{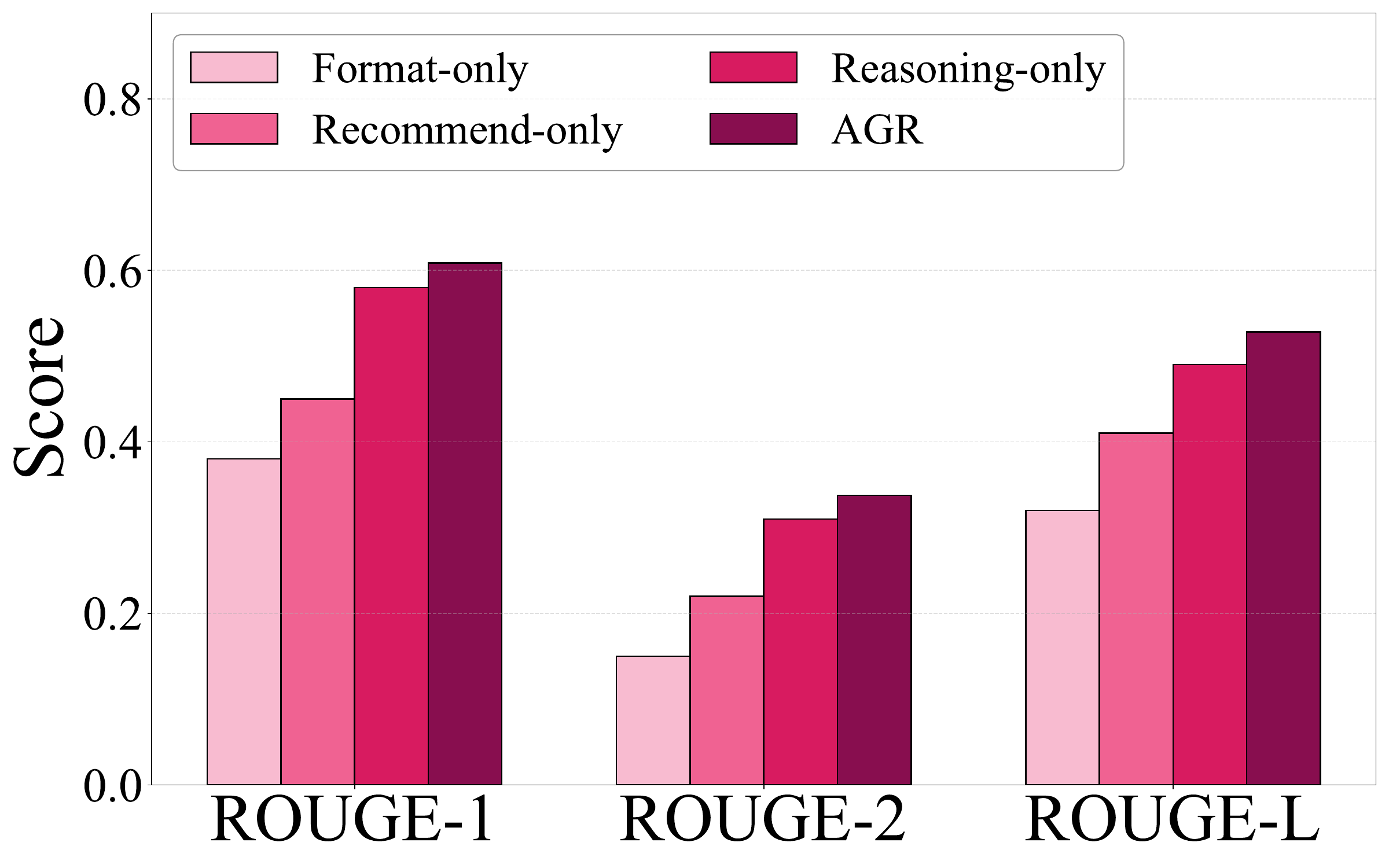}}	
	\subfigure[Human Evaluation Scores]{\label{subfig:reward3.pdf}
		\includegraphics[width=0.21\textwidth]{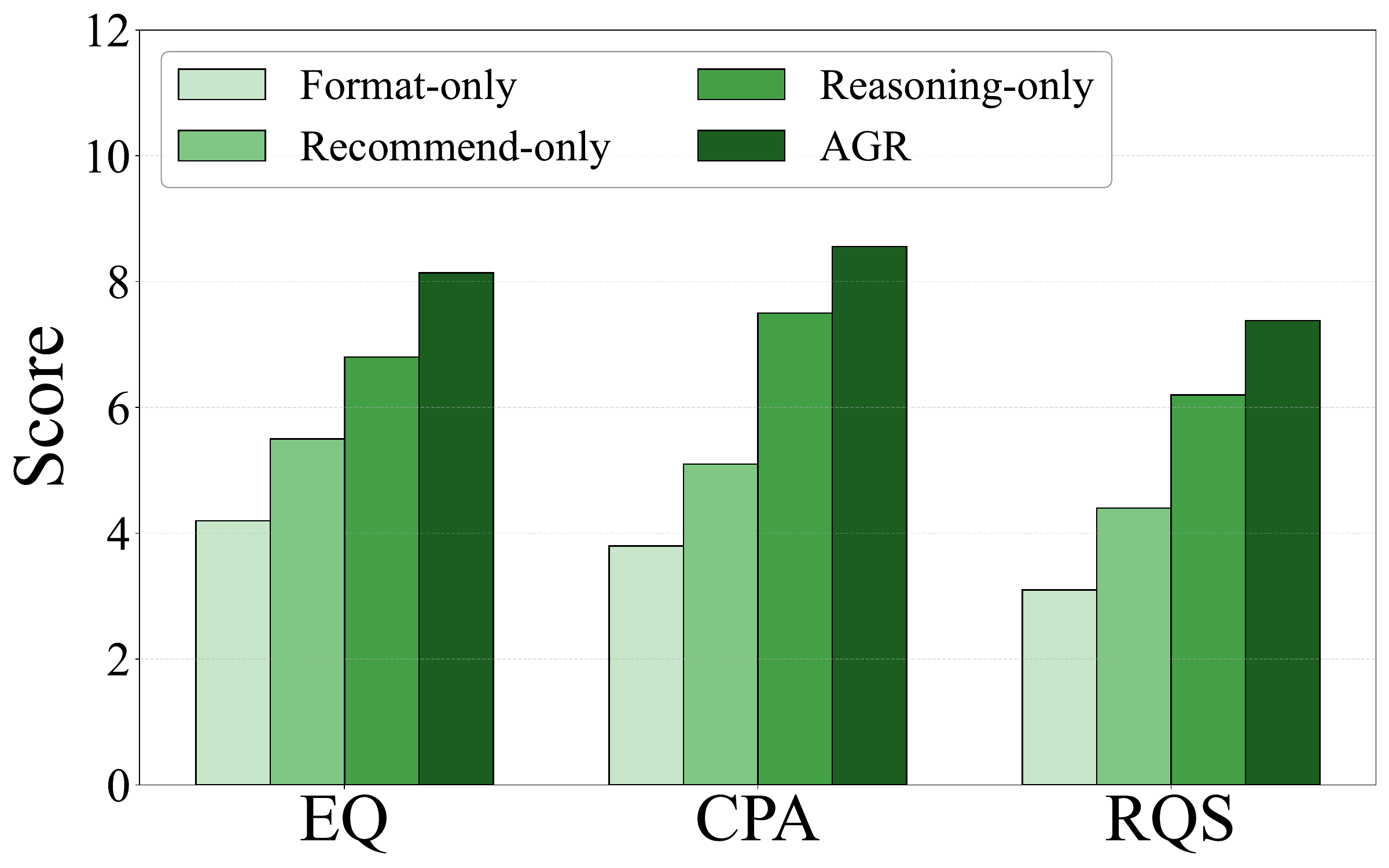}}	
	\caption{Ablation study of reward function combinations, $p < 0.01$.}
	\label{fig:reward}
	\vspace*{-8pt}
\end{figure}

We conduct ablation studies to evaluate three key components: the Memory and Reasoning Modules (RQ3), the SFT and GRPO training stages (RQ4), and individual reward functions (RQ5). For RQ3 and RQ4, we report the results in Tables~\ref{tb:overall_lastfm_t}–\ref{tb:overall_douban_t}. The results show that: 1) Memory Module is essential; without it (AGR-R), recommendation scores drop, showing that it tracks evolving user/group interests for robust recommendations.
2) Reasoning Module is necessary; its removal (AGR-M) lowers explainability metrics, confirming it explicitly models group decision-making.
3) Both SFT and GRPO are required in RFT; SFT is a prerequisite for GRPO, and ablation studies confirm that neither component alone (AGR-SFT or AGR-GRPO) achieves the performance of the complete AGR framework.
4) We further examine the judgement bias in LLM evaluation within the Reasoning Module. Specifically, we compare our proposed \textit{ICL+LLM-as-a-judge} approach with two variants: a) \textit{Pure LLM-as-a-judge}: directly scoring by LLM without any human annotation; and
b) \textit{LLM-as-a-judge + Human Annotation}: human scoring based on LLM-generated outputs. We report Fleiss' Kappa~\cite{fleiss1971measuring} between human consensus and LLM scores, and HR@5 for ranking quality in Fig.~\ref{fig:ablation_lms}. Results show that \textit{ICL+LLM-as-a-judge} consistently outperforms both variants on both metrics, demonstrating its effectiveness in reducing LLM scoring bias.
5) To explore robustness against prompt variations, we conduct a prompt ablation study on 200 samples from LastFM using two variants: a) synonym replacement with slight rephrasing, and b) partial removal of evaluation dimensions. Table~\ref{tb:prompt_ablation} shows AGR is insensitive to prompt wording, with BERTScore$>$0.92 and ROUGE*$>$0.81 across all variants.
6) For RQ5, we assess each GRPO reward component and report the results in Fig.~\ref{fig:reward}. The results show that: all GRPO reward modules are necessary. Removing format reward reduces overall performance; removing recommendation or reasoning reward degrades recommendation or explainability scores, respectively.

\hide{
	(7)To further verify the contrastive learning mechanism of GRPO, we fix the total number of rollouts to 4 and systematically adjust the ratio of positive samples \(m\) and negative samples \(n\) (where \(m + n = 4\) and \(m \neq n\)) to explore the impact of imbalanced positive–negative samples on model performance.
	We set three typical contrastive settings: \((m=1, n=3)\), \((m=2, n=2)\), and \((m=3, n=1)\), corresponding to \textit{positive-minority, balanced, and negative-minority} scenarios, respectively.
	As shown in Fig.~\ref{fig:mn_ratio}, the experimental results show that:
	1) As long as valid positive–negative contrast signals exist, the model can achieve near-optimal performance, indicating that the optimization core of GRPO relies on the \textit{relative advantage} between positive and negative samples rather than strict quantitative balance.
	2) The asymmetric setting \(m \neq n\) still enables stable learning, proving that GRPO does not rely on symmetric positive–negative samples and further supporting its nature as contrastive policy optimization.
	3) When the number of positive samples is slightly larger than that of negative samples (\(m > n\)), the model achieves small but consistent performance gains, which aligns with the task characteristic that positive signals are more decisive in personalized recommendation.
}


\subsection{Model Efficiency Analysis (RQ6)}
To assess the efficiency of the Memory Module's core operations and each stage within the Reasoning Module (RQ6), we decompose AGR's inference pipeline into multiple stages for cost analysis. As shown in Table~\ref{tb:unified_efficiency}, we measure the processing times of the Memory Module's core operations (insertion, updating, forgetting, retrieval, and summarization) and the processing times of each stage in the Reasoning Module. Additionally, we also report input/output token counts, time complexity, and relative time proportion for each stage on the LastFM dataset. The results show that the Memory Module and Reasoning Module add only negligible overhead to the execution time due to their low time complexity.

\begin{table}[t]
	\centering
	\caption{Unified efficiency analysis of memory operations, reasoning stages, and model-level comparison.}
	\label{tb:unified_efficiency}
	\begin{tabular}{lccc}
		\toprule
		\textbf{Stage} & \textbf{Time (s)} & \textbf{Tokens} & \textbf{Time Proportion}  \\
		\midrule
		\multicolumn{4}{l}{\textbf{Memory Module}} \\
		\midrule
		Insertion & 0.0023 & -- & 0.18\%  \\
		Forgetting & 0.0016 & -- & 0.12\%  \\
		Updating & 0.0097 & -- & 0.75\%  \\
		Retrieval & 0.0217 & -- & 1.69\%  \\
		Summarization & 0.2517 & 200 & 19.55\%  \\
		\midrule
		\multicolumn{4}{l}{\textbf{Reasoning Module}} \\
		\midrule
		Interests Collection & 0.0699 & 53 & 5.43\%  \\
		Consensus Refinement & 0.0974 & 75 & 7.56\% \\
		Multi-dim Evaluation & 0.3635 & 131 & 28.22\%  \\
		Generation & 0.4701 & 971.68 & 36.51\%  \\
		\midrule
		\multicolumn{4}{l}{\textbf{Comparison Models}} \\
		\midrule
		LLaMA (Vanilla) & 0.7163 & 1,124.2 & --  \\
		LLaMA+CoT+ICL & 0.9827 & 1,064.5 & --  \\
		AGR (overall) & 1.2879 & 971.68 & --  \\
		\bottomrule
	\end{tabular}
	\vspace{-16pt}
\end{table}

\hide{\begin{figure}[h]
		\centering
		\subfigure[BLEU \& ROUGE*]{\label{subfig:rollout1}
			\includegraphics[width=0.21\textwidth]{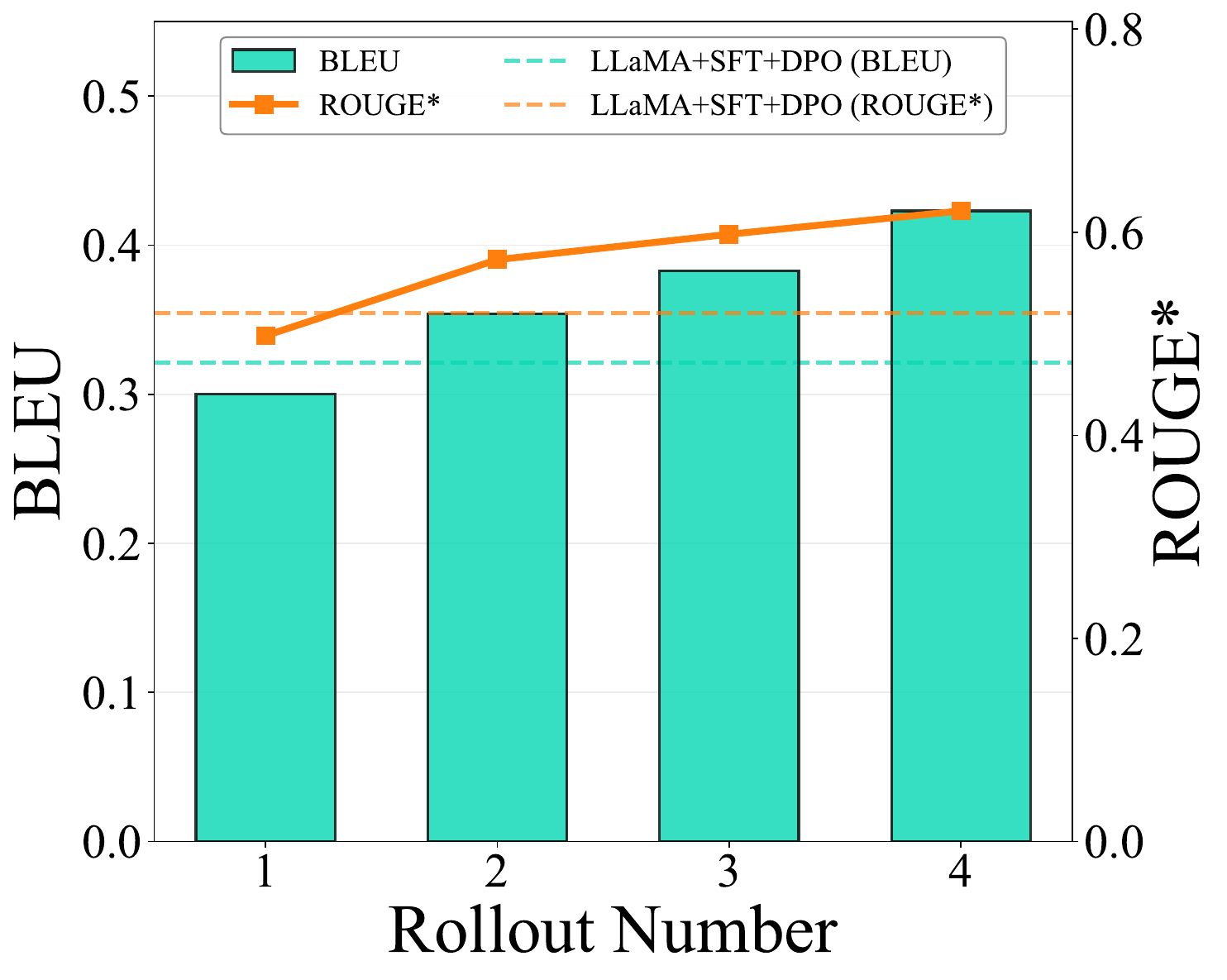}}
		\subfigure[NDCG@5 \& HR@5]{\label{subfig:rollout2}
			\includegraphics[width=0.21\textwidth]{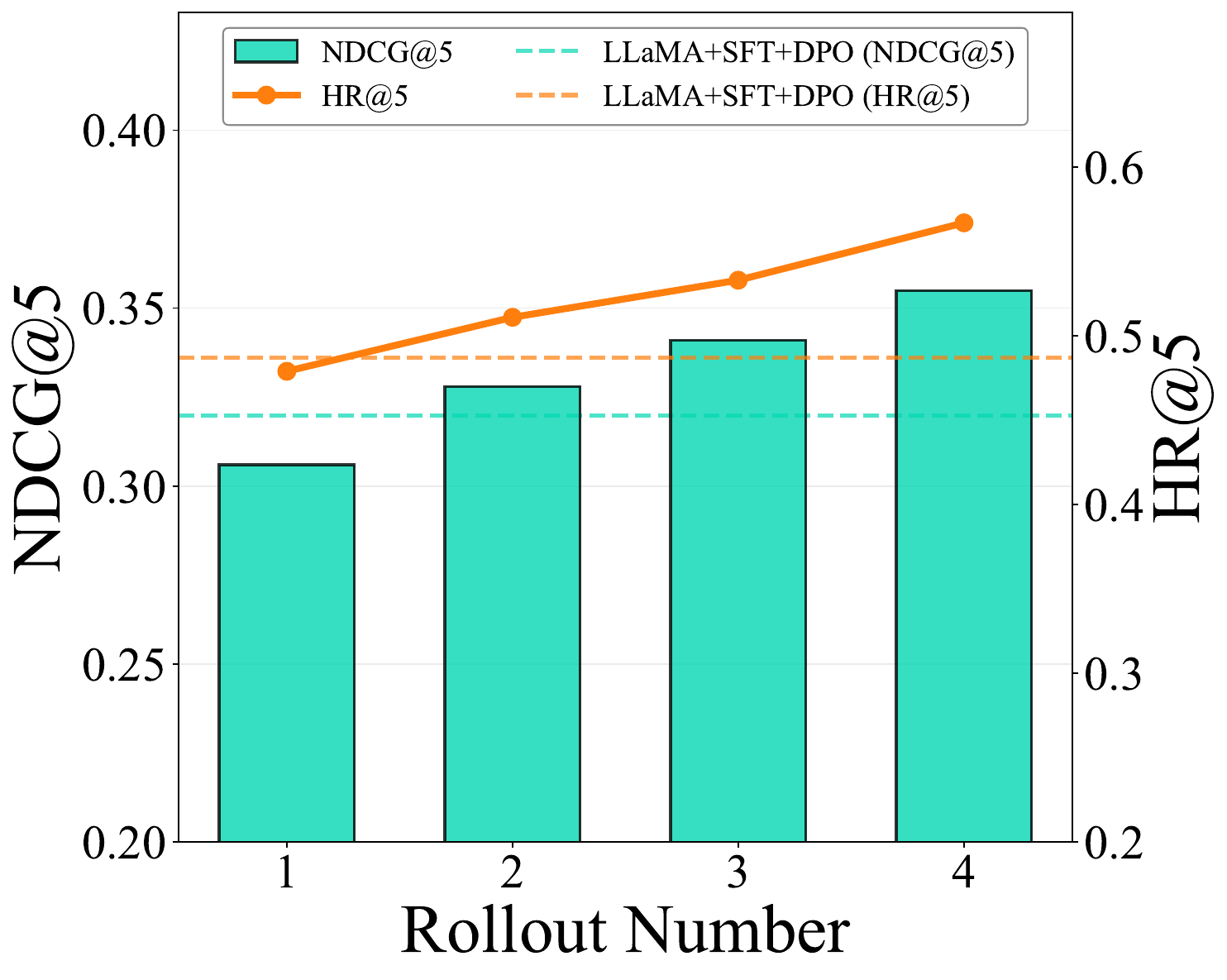}}	
		\caption{Comparison of different rollout numbers.}
		\label{fig:rollout}
\end{figure}}

\subsection{Hyper-parameter Analysis (RQ7)}
\begin{figure}[h]
		\centering
		\subfigure[HR@5]{\label{subfig:rq6_forget_hr5}
			\includegraphics[width=0.46\linewidth]{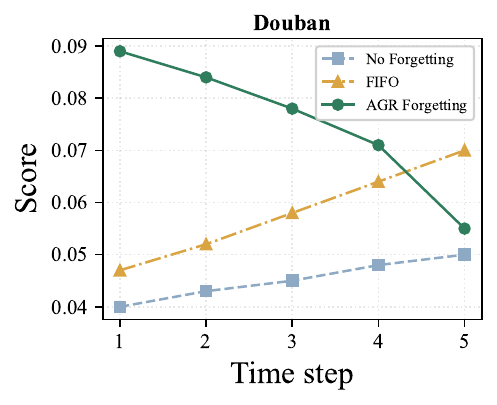}}
		\hfill
		\subfigure[BLEU]{\label{subfig:rq6_forget_bleu}
			\includegraphics[width=0.46\linewidth]{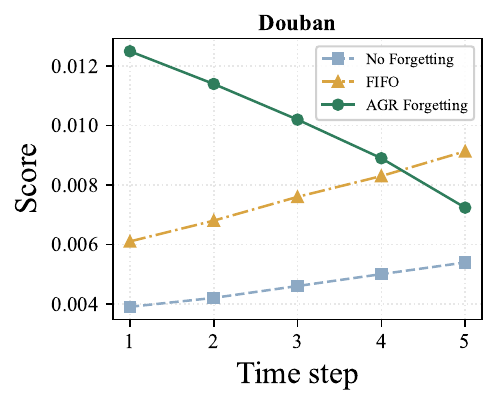}}
		\subfigure[MRR]{\label{subfig:rq6_forget_mrr}
			\includegraphics[width=0.46\linewidth]{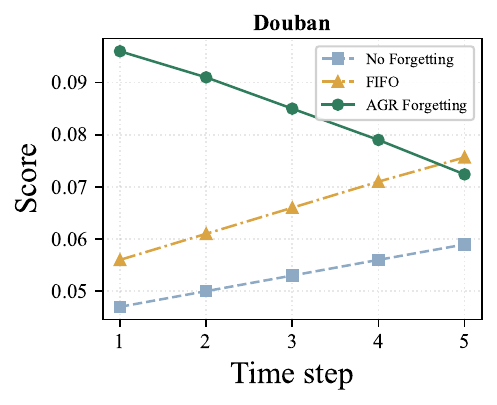}}
		\hfill
		\subfigure[CPA]{\label{subfig:rq6_forget_cpa}
			\includegraphics[width=0.46\linewidth]{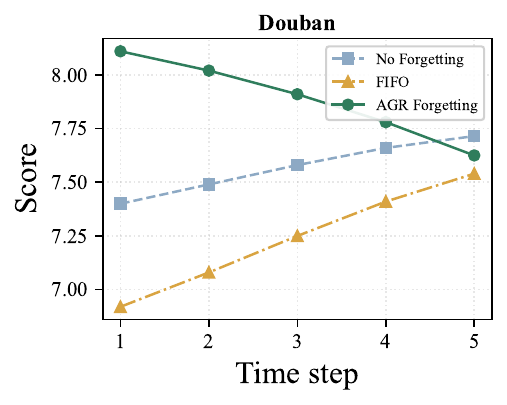}}
		\caption{Forgetting comparison of irrelevant forgetting mechanism and its variant on Douban across five time steps.}
		\label{fig:rq6_forgetting}
	\end{figure}

We investigate the impact of varying hyper-parameter configurations on model performance.\hide{ 1) Specifically, we use LLaMA3-8B as the backbone model, and examine the influence of the rollout number within the GRPO framework on the LastFM dataset. The results are presented in Fig.~\ref{fig:rollout}. All experiments are repeated five times to obtain the mean values and variances. We find that when four or more rollouts are used, the performance of SFT+GRPO surpasses that of SFT+DPO. Under identical experimental settings, the average variance of the GRPO method is approximately \textbf{7.9e-5}, which is much lower than \textbf{3.16e-4} of the DPO method, demonstrating better convergence and stability. }
1) We analyze the effect of memory capacity \(\varphi\) on performance. As shown in Table~\ref{tb:phi_ablation}, the best performance occurs at \(\varphi = 8\) (HR@5 = 0.70). A small capacity (\(\varphi = 2\)) leads to early forgetting of key information, while a large capacity (\(\varphi = 20\)) adds noise and computational cost. Thus, we set \(\varphi = 8\) for all experiments.
2) To evaluate whether the proposed irrelevant forgetting mechanism (AGR Forgetting) can effectively capture evolving group/user profiles, we adopt the \textit{leave-last-t-out} strategy and compare two forgetting variants: a) Retain ALL (No Forgetting); and b) Chronological Forgetting (FIFO), which discards the oldest entries. For AGR Forgetting and FIFO, the Memory capacity is set as $\varphi=8$. We randomly select 500 Douban samples, each with an average of 10.8 member interactions per group. Fig.~\ref{fig:rq6_forgetting} shows that: 1) when $t$ is large ($t$=5, the data sparsity settings), AGR underperforms FIFO due to sparse interactions.  2) when $t$ decreases ($t$=1,2,3,4), AGR performs the best, showing it selectively removes irrelevant items for long-term memory. 3) The Retain ALL strategy performs the worst, as it cannot capture the group profile accurately.

\hide{
	\textcolor{green}{We conduct ablation studies on three key hyperparameters in the dynamic trigger mechanism, including similarity threshold $\theta_{\rm sim}$, interaction threshold $\theta$, and time interval $\tau$. As shown in Fig.~\ref{fig:trigger_ablation}, our settings ($\theta_{\rm sim}=0.4$, $\theta=7$, $\tau=3$) achieve the optimal HR@5 performance, demonstrating the rationality and robustness of our parameter selection.}

	\begin{figure}[h]
		\centering
		\subfigure[$\theta_{\rm sim}$]{\label{subfig:theta_sim}
			\includegraphics[width=0.31\linewidth]{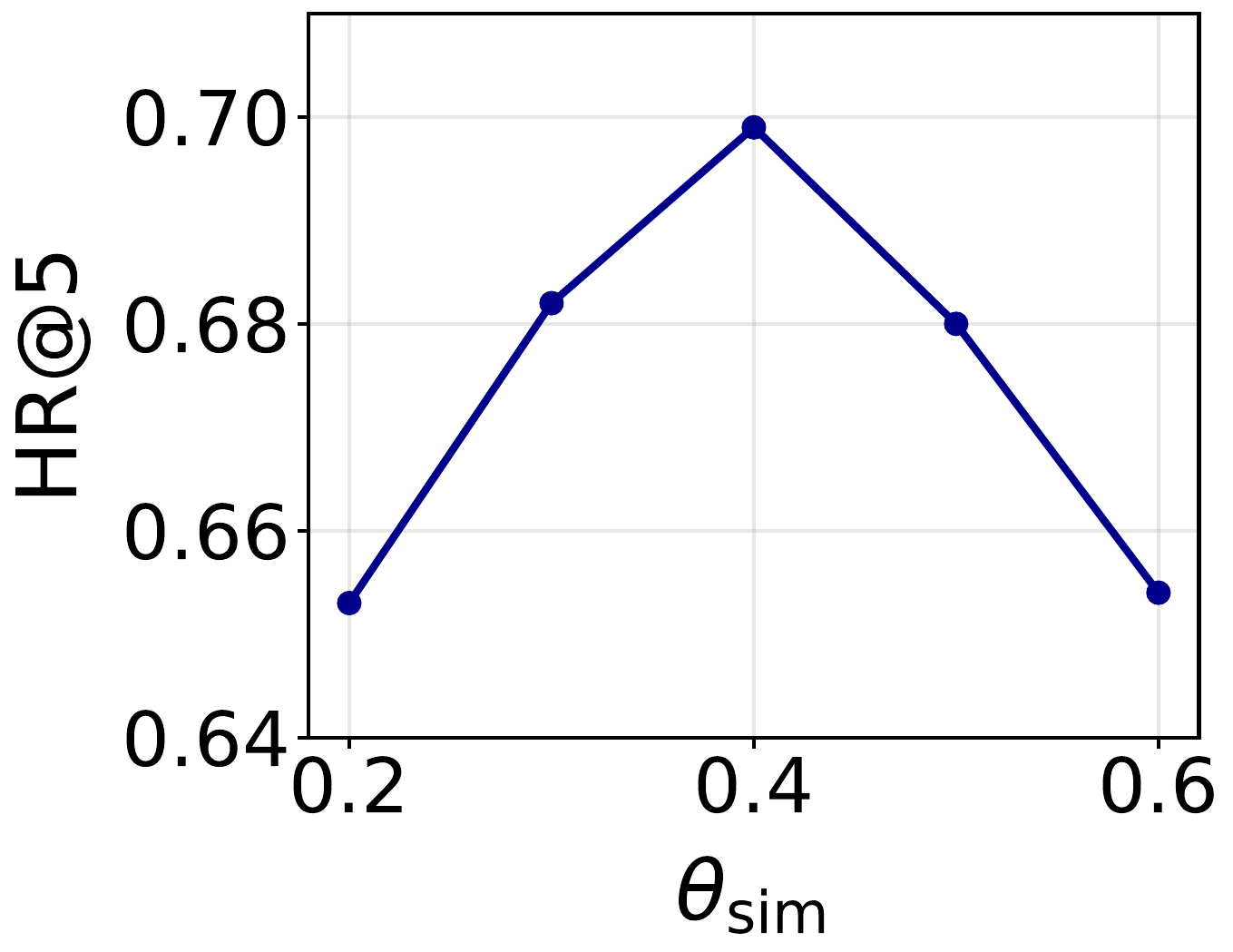}}
		\subfigure[$\theta$]{\label{subfig:theta}
			\includegraphics[width=0.31\linewidth]{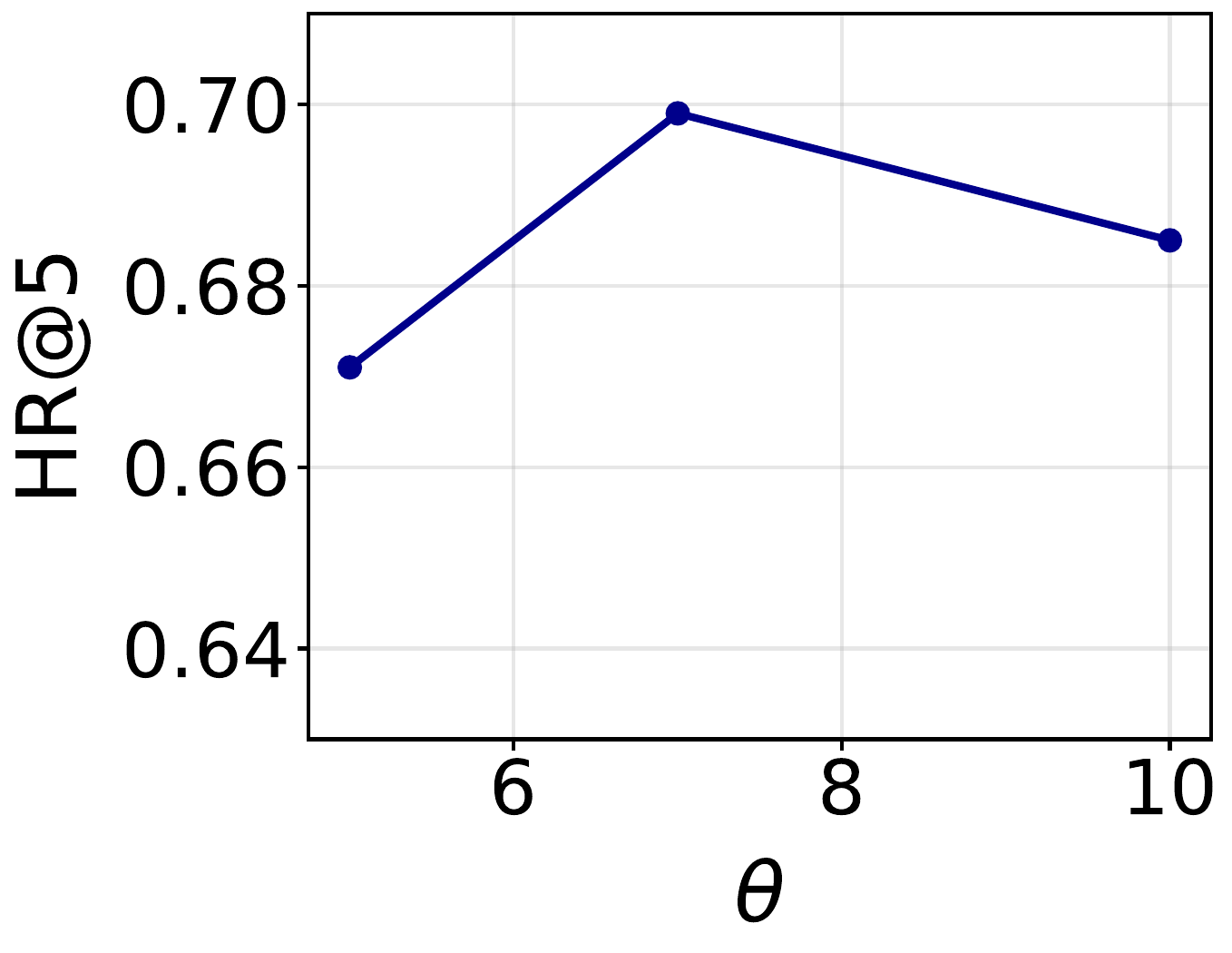}}
		\subfigure[$\tau$]{\label{subfig:tau}
			\includegraphics[width=0.31\linewidth]{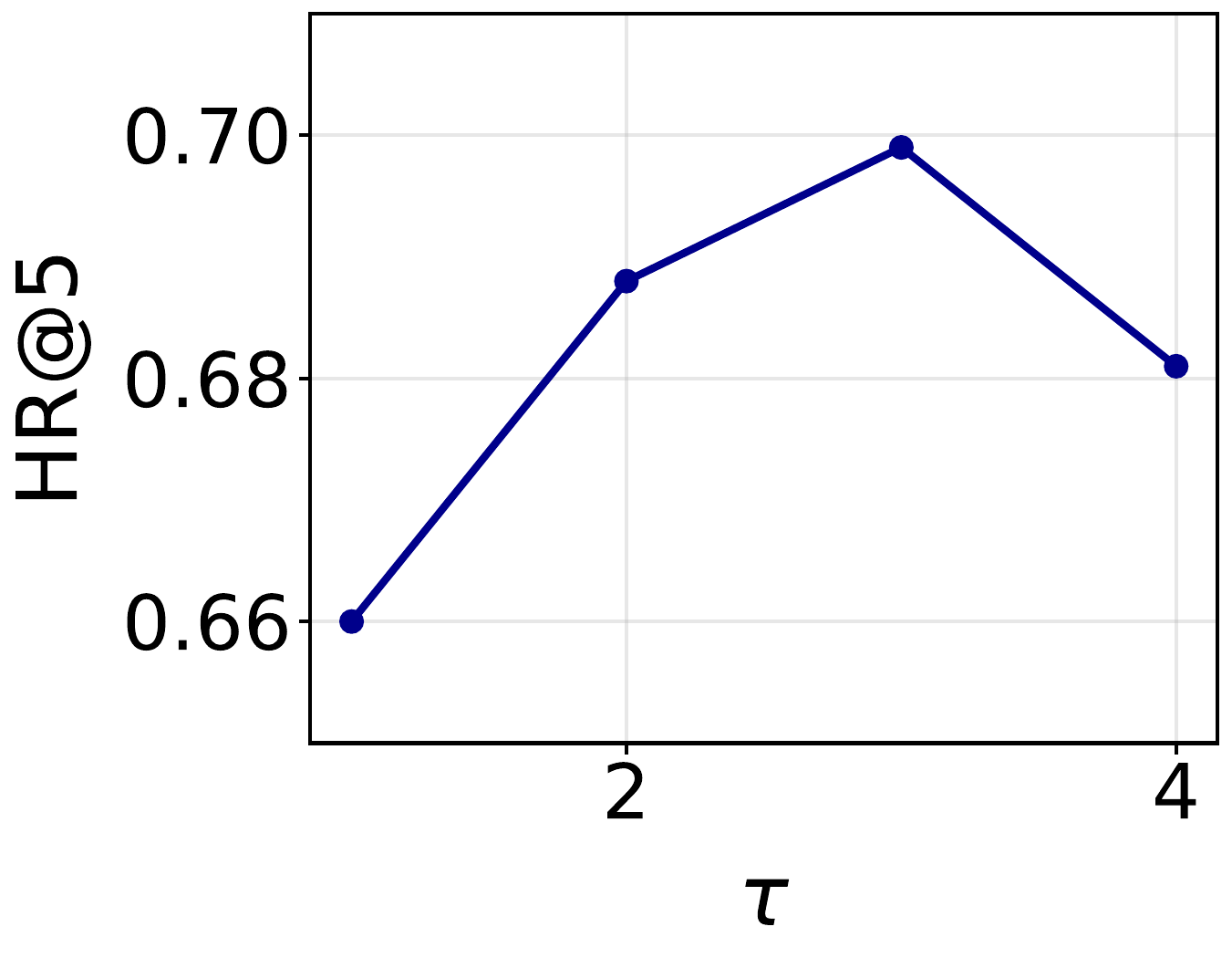}}
		\caption{Ablation studies on hyperparameters of the dynamic trigger mechanism.}
		\label{fig:trigger_ablation}
	\end{figure}
}

\begin{table}[htbp]
	\vspace{-16pt}
	\centering
	\caption{Ablation study of memory capacity \(\varphi\) on LastFM.}
	\setlength{\tabcolsep}{7pt}  
	\begin{tabular}{ccccccc}
		\toprule
		\(\varphi\) & HR@5 & NDCG@5 & CPA & PDA & RQS & EQ \\
		\midrule
		2  & 0.65 & 0.52 & 7.90 & 7.71 & 6.96 & 7.89 \\
		4 & 0.69 & 0.55 & 8.39 & 8.18 & 7.39 & 8.38 \\
		8 & \textbf{0.70} & \textbf{0.56} & \textbf{8.51} & \textbf{8.30} & \textbf{7.50} & \textbf{8.50} \\
		16 & 0.67 & 0.53 & 8.15 & 7.94 & 7.18 & 8.14 \\
		20 & 0.64 & 0.50 & 7.78 & 7.59 & 6.86 & 7.77 \\
		\bottomrule
	\end{tabular}
	\label{tb:phi_ablation}
\end{table}

	\section{Conclusion}
We propose AGR, a LLM-based Agent with a Memory Module and a Reasoning Module. The Memory Module uses a token-based hash table to dynamically manage the historical interactions of groups and users. The Reasoning Module then performs multi-step reasoning on profiles retrieved from the Memory Module. AGR is trained using RFT, which enables the agent to autonomously invoke its Memory and Reasoning Modules. Experiments on two datasets demonstrate that AGR significantly outperforms state-of-the-art methods in both recommendation accuracy and explainability.
	\section*{Acknowledgment}
This research was supported by the National Natural Science Foundation of China (No. 62402328), the Undergraduate “Qiyan” Program of Beijing Natural Science Foundation (No. QY25392),  the Capital Normal University Talent Support Program (No.26563005079), the Australian Research Council partially supports this work under the streams of Future Fellowship (Grant No. FT210100624), the Discovery Project (Grant No. DP260100326), and the Linkage Project (Grant No. LP230200892, LP240200546 and LP250200778).
	
	\normalem
	\bibliographystyle{IEEEtran}
	\bibliography{IEEEabrv,myrefs}

\begin{thebibliography}{10}
\providecommand{\url}[1]{#1}
\csname url@samestyle\endcsname
\providecommand{\newblock}{\relax}
\providecommand{\bibinfo}[2]{#2}
\providecommand{\BIBentrySTDinterwordspacing}{\spaceskip=0pt\relax}
\providecommand{\BIBentryALTinterwordstretchfactor}{4}
\providecommand{\BIBentryALTinterwordspacing}{\spaceskip=\fontdimen2\font plus
\BIBentryALTinterwordstretchfactor\fontdimen3\font minus
  \fontdimen4\font\relax}
\providecommand{\BIBforeignlanguage}[2]{{%
\expandafter\ifx\csname l@#1\endcsname\relax
\typeout{** WARNING: IEEEtran.bst: No hyphenation pattern has been}%
\typeout{** loaded for the language `#1'. Using the pattern for}%
\typeout{** the default language instead.}%
\else
\language=\csname l@#1\endcsname
\fi
#2}}
\providecommand{\BIBdecl}{\relax}
\BIBdecl

\bibitem{DBLP:conf/icde/YinW0LYZ19}
H.~Yin, Q.~Wang, and K.~Z. et~al, ``Social influence-based group representation
  learning for group recommendation,'' in \emph{ICDE'19}.

\bibitem{DBLP:conf/recsys/BaltrunasMR10}
L.~Baltrunas, T.~Makcinskas, and F.~Ricci, ``Group recommendations with rank
  aggregation and collaborative filtering,'' in \emph{RecSys'10}.

\bibitem{DBLP:journals/pvldb/Amer-YahiaRCDY09}
S.~Amer{-}Yahia, S.~B. Roy, and A.~C. et~al, ``Group recommendation: Semantics
  and efficiency,'' \emph{Proc. {VLDB} Endow.}, 2009.

\bibitem{DBLP:series/sci/BorattoC11}
L.~Boratto and S.~Carta, ``State-of-the-art in group recommendation and new
  approaches for automatic identification of groups,'' in \emph{Information
  Retrieval and Mining in Distributed Environments}, 2011.

\bibitem{DBLP:conf/intrs/LubosTLGHWF25}
S.~Lubos, T.~N.~T. Tran, and V.~L. et~al, ``Towards llm-enhanced group
  recommender systems,'' in \emph{RecSys'25}.

\bibitem{DBLP:conf/ictai/LubosFGL25}
S.~Lubos, A.~Felfernig, and D.~G. et~al, ``Assessing llms for prioritization in
  meeting-based group recommendations,'' in \emph{ICTAI'25}.

\bibitem{DBLP:conf/um/LubosFGLHWF25}
\relax Sebastian~Lubos, A.~Felfernig, and D.~G. et~al, ``Towards group decision
  support with llm-based meeting analysis,'' in \emph{UMAP'25}.

\bibitem{DBLP:journals/tmlr/ZhangGYYZTZLXLZCZFWHVLW26}
G.~Zhang, H.~Geng, and X.~Y. et~al, ``The landscape of agentic reinforcement
  learning for llms: {A} survey,'' \emph{Trans. Mach. Learn. Res.}, 2026.

\bibitem{DBLP:conf/nips/SchickDDRLHZCS23}
T.~Schick, J.~Dwivedi{-}Yu, and R.~D. et~al, ``Toolformer: Language models can
  teach themselves to use tools,'' in \emph{{NeurIPS'23}}.

\bibitem{DBLP:conf/kdd/YuanCL14}
Q.~Yuan, G.~Cong, and C.~Lin, ``{COM:} a generative model for group
  recommendation,'' in \emph{{KDD'14}}.

\bibitem{DBLP:journals/neco/HochreiterS97}
S.~Hochreiter and J.~Schmidhuber, ``Long short-term memory,'' \emph{Neural
  Comput.}, 1997.

\bibitem{DBLP:conf/ssst/ChoMBB14}
K.~Cho, B.~van Merrienboer, and D.~B. et~al, ``On the properties of neural
  machine translation: Encoder-decoder approaches,'' in \emph{EMNLP'14
  Workshop}.

\bibitem{DBLP:journals/corr/abs-2411-09425}
X.~Lv, J.~Cao, and S.~G. et~al, ``{MARM:} unlocking the future of
  recommendation systems through memory augmentation and scalable complexity,''
  \emph{CoRR}, 2024.

\bibitem{DBLP:conf/www/LuCZCXXZW25}
H.~Lu, Z.~Chai, and Y.~Z. et~al, ``Large memory network for recommendation,''
  in \emph{{WWW} 2025}.

\bibitem{DBLP:conf/cikm/XiLL0T0024}
Y.~Xi, W.~Liu, and J.~L. al, ``Memocrs: Memory-enhanced sequential
  conversational recommender systems with large language models,'' in
  \emph{CIKM'24}.

\bibitem{DBLP:journals/corr/abs-2510-14629}
J.~Huang, X.~Zou, and L.~X. et~al, ``Mr.rec: Synergizing memory and reasoning
  for personalized recommendation assistant with llms,'' \emph{CoRR}, 2025.

\bibitem{DBLP:conf/kdd/Wang00LC19}
X.~Wang, X.~He, and Y.~C. et~al, ``{KGAT:} knowledge graph attention network
  for recommendation,'' in \emph{KDD'19}.

\bibitem{DBLP:conf/www/MaZCJWLMR19}
W.~Ma, M.~Zhang, and Y.~C. et~al, ``Jointly learning explainable rules for
  recommendation with knowledge graph,'' in \emph{WWW'19}.

\bibitem{DBLP:conf/icdm/KangM18}
W.~Kang and J.~J. McAuley, ``Self-attentive sequential recommendation,'' in
  \emph{ICDM'18}.

\bibitem{DBLP:conf/aaai/YueYZSLW25}
W.~Yue, Y.~Yin, and X.~Z. et~al, ``Cot4rec: Revealing user preferences through
  chain of thought for recommender systems,'' in \emph{AAAI'25}.

\bibitem{DBLP:conf/kdd/LeiLYHL024}
Y.~Lei, J.~Lian, and J.~Y. et~al, ``Recexplainer: Aligning large language
  models for explaining recommendation models,'' in \emph{KDD'24}.

\bibitem{DBLP:journals/corr/abs-2505-09388}
Q.~Team, ``Qwen3 technical report,'' \emph{CoRR}, 2025.

\bibitem{8594636}
Y.~A. Malkov and D.~A. Yashunin, ``Efficient and robust approximate nearest
  neighbor search using hierarchical navigable small world graphs,''
  \emph{TPAMI'20}.

\bibitem{DBLP:journals/corr/abs-2411-15594}
J.~Gu, X.~Jiang, and Z.~S. et~al, ``A survey on llm-as-a-judge,'' \emph{CoRR},
  2024.

\bibitem{DBLP:conf/nips/MadaanTGHGW0DPY23}
A.~Madaan, N.~Tandon, and P.~G. et~al, ``Self-refine: Iterative refinement with
  self-feedback,'' in \emph{NeurIPS'23}.

\bibitem{DBLP:conf/nips/BrownMRSKDNSSAG20}
T.~B. Brown, B.~Mann, and N.~R. et~al, ``Language models are few-shot
  learners,'' in \emph{NeurIPS'20}.

\bibitem{DBLP:journals/corr/abs-2501-12948}
DeepSeek{-}AI, ``Deepseek-r1: Incentivizing reasoning capability in llms via
  reinforcement learning,'' \emph{CoRR}, 2025.

\bibitem{DBLP:journals/corr/abs-2402-03300}
Z.~Shao, P.~Wang, and Q.~Z. et~al, ``Deepseekmath: Pushing the limits of
  mathematical reasoning in open language models,'' \emph{CoRR}, 2024.

\bibitem{DBLP:journals/is/KimE15}
H.~Kim and A.~E. Saddik, ``A stochastic approach to group recommendations in
  social media systems,'' \emph{Inf. Syst.}, 2015.

\bibitem{DBLP:journals/ijdsa/ZhengLSZLW17}
J.~Zheng, J.~Liu, and C.~S. et~al, ``Recommendation in heterogeneous
  information network via dual similarity regularization,'' \emph{Int. J. Data
  Sci. Anal.}, 2017.

\bibitem{DBLP:conf/sigir/Cao0MAYH18}
D.~Cao, X.~He, and L.~M. et~al, ``Attentive group recommendation,'' in
  \emph{SIGIR'18}.

\bibitem{DBLP:conf/aaai/Ye0WCZ025}
G.~Ye, W.~Wu, and G.~W. et~al, ``Disentangled modeling of preferences and
  social influence for group recommendation,'' in \emph{AAAI'25}.

\bibitem{DBLP:journals/tois/Tian0SJZ22}
Z.~Tian, Y.~Liu, and J.~S. et~al, ``Exploiting group information for
  personalized recommendation with graph neural networks,'' \emph{TOIS'22}.

\bibitem{DBLP:conf/aciids/DobrovolnySK21}
M.~Dobrovolny, A.~Selamat, and O.~Krejcar, ``Session based recommendations
  using recurrent neural networks - long short-term memory,'' in
  \emph{ACIIDS'21}.

\bibitem{DBLP:conf/emnlp/XieWJYM25}
J.~Xie, H.~Wu, and D.~J. et~al, ``Lohrec: Leveraging order and hierarchy in
  generative sequential recommendation,'' in \emph{EMNLP'25 Findings}.

\bibitem{DBLP:journals/corr/abs-2407-21783}
L.~Team, ``The llama 3 herd of models,'' \emph{CoRR}, 2024.

\bibitem{DBLP:conf/acl/KothapalliFS25}
V.~Kothapalli, H.~Firooz, and M.~Sanjabi, ``Cot-icl lab: {A} synthetic
  framework for studying chain-of-thought learning from in-context
  demonstrations,'' in \emph{ACL'25}.

\bibitem{fleiss1971measuring}
J.~L. Fleiss, ``Measuring nominal scale agreement among many raters.''
  \emph{Psychological bulletin}, 1971.

\bibitem{Mann1947OnAT}
H.~B. Mann and D.~R. Whitney, ``On a test of whether one of two random
  variables is stochastically larger than the other,'' \emph{Annals of
  Mathematical Statistics}, 1947.

\end{thebibliography}
	
	\vspace{12pt}
	
\end{document}